\documentclass[10pt,a4paper]{article}
\usepackage[utf8]{inputenc}
\usepackage{amsmath}
\usepackage{amsfonts}
\usepackage{amssymb}
\usepackage{color}
\usepackage{graphicx}
\usepackage{float}
\usepackage[square]{natbib}
\usepackage{doi}
\hypersetup{hidelinks}
\usepackage[a4paper, total={6in, 8in}]{geometry}
\usepackage{tikz}
\usepackage[export]{adjustbox}
\usepackage{layouts}
\usepackage{psfrag}
\usetikzlibrary{shapes.geometric, arrows}

\usepackage[hang,flushmargin,symbol]{footmisc}
\usepackage{subcaption}
\usepackage{tikz,pgfplots}
\usepackage{siunitx}
\usepackage{mathtools}
\usepackage{definitions}

\begin{document}

\begin{center}
\textbf{\Large{Rate-Independent Gradient Crystal Plasticity Theory\\ 
		Robust Algorithmic Formulations
		based on Incremental Energy Minimization}}
~\\
~\\
~\\
Volker Fohrmeister and J\"orn Mosler\footnote{\noindent Corresponding author\\Email adress: joern.mosler@tu-dortmund.de}
~\\
~\\
\textit{Institute of Mechanics, TU Dortmund, Leonhard-Euler-Str. 5, 44227 Dortmund, Germany}
~\\
~\\
\end{center}
Abstract:
Numerically robust algorithmic formulations suitable for rate-independent crystal plasticity are presented. They cover classic local models as well as gradient-enhanced theories in which the gradients of the plastic slips are incorporated by means of the micromorphic approach. The elaborated algorithmic formulations rely on the underlying variational structure of (associative) crystal plasticity. To be more precise and in line with so-called variational constitutive updates or incremental energy minimization principles, an incrementally defined energy derived from the underlying time-continuous constitutive model represents the starting point of the novel numerically robust algorithmic formulations. This incrementally defined potential allows to compute all variables jointly as minimizers of this energy. While such discrete variational constitutive updates are not new in general, they are considered here in order to employ powerful techniques from non-linear constrained optimization theory in order to compute robustly the aforementioned minimizers. The analyzed prototype models are based on (1) nonlinear complementarity problem (NCP) functions as well as on (2) the augmented Lagrangian formulation. Numerical experiments show the numerical robustness of the resulting algorithmic formulations. Furthermore, it is shown that the novel algorithmic ideas can also be integrated into classic, non-variational, return-mapping schemes. 
~
\thispagestyle{empty}
\clearpage

\section{INTRODUCTION}\label{sec:intro}

Although crystal plasticity theory dates back to the early pioneering works \cite{Taylor1938,Bishop_Hill:51a,Bishop_Hill:51b} and thus, it was already proposed more than eighty years ago, it still represents a sound and powerful framework for understanding the mechanical response of (single) crystals. Furthermore, despite its simple structure and despite the relatively simple underlying assumptions, crystal plasticity theory allows to capture very complex phenomena such as texture evolution in polycrystals, cf. \cite{Miehe2002,Homayonifar2012}. For this reason, crystal plasticity theory is still frequently and successfully applied to simulate crystal behaviour, see \cite{MA20062181,SISKA2009793,AMCuitino_1993,Homayonifar2011,Kochmann2022,Daroju2022}.

While the structure of crystal plasticity is indeed relatively simple (but still physically sound), the algorithmic formulation is very challenging. To be more precise and as far as rate-independent local crystal plasticity is concerned, the computation of the active slip systems is not unique and results in severe numerical problems. The well-known reason for this non-uniqueness is that the space of stress deviators (5 for a symmetric stress tensor, see e.g. \cite{Hill_Rice:1972,Bassani_Wu:1991,Havner:1992,FohrmeisterDiazMosler_2018a}) is smaller than the number of possibly active slip systems (e.g., 12 for an FCC crystal). Clearly, one might argue that real crystals also show a certain viscosity which can regularize the aforementioned non-uniqueness. However, this beneficial effect is not present for very slow processes and furthermore, combining or mixing a physics-motivated viscosity with a numerically-motivated one is problematic, since both effects cannot be separated in general and hence, they interfere, cf. \cite{LangenfeldJunkerMosler_2018}. For this reason, focus is on a rate-independent setting.

In order to control the non-uniqueness of crystal plasticity theory and in order to also identify the active slip systems within the algorithmic treatment, several ideas have been proposed. The probably most prominent ones are the application of a pseudo inverse (see, e.g. \cite{ANAND1996525,SCHRODER1997168,Manik2014}), the introduction of a numerical viscosity (see, e.g. \cite{MieheSchroeder_2001,Manik_Asadkandi_Holmedal:2022,Rys2022}) or the active set method (see, e.g. \cite{Simo1998,SimoHughes1998,MieheSchroeder_2001}). Other algorithms are based on the postulate of maximum dissipation. Focusing on associative evolution equations, crystal plasticity theory can, for instance, be re-written as such an optimization problem. This opens up the possibility of applying powerful methods from non-linear (constraint) optimization. This line of thought was followed in \cite{Schmidt-Baldassari:03}, in which two algorithms were proposed. The first of those falls into the range of Nonlinear complementarity problem (NCP) functions, while the second one adopts the Augmented Lagrangian framework. More recently, the interior point method --- another non-linear (constraint) optimization algorithm -- has also been applied to crystal plasticity theory, cf. \cite{SCHEUNEMANN20201,SCHEUNEMANN2021111149,Niehueser_Mosler:23}.

Within the present paper, two non-linear (constraint) optimization algorithms are proposed in order to (1) identify the active slip systems and (2) in order to compute the unknown plastic slips: A method based on nonlinear complementarity problem (NCP) functions, and an algorithm relying on the Augmented Lagrangian framework. However and in contrast to the papers cited before such as \cite{Schmidt-Baldassari:03} where the postulate of maximum dissipation is considered, a global variational principle represents the starting point here. To be more precise and in line with \cite{Ortiz_Stainier:99,Ortiz_Repetto:99,Carstensen_Hackl_Mielke:02,Miehe:02,Mosler-Bruhns2010,Stainier2010,FohrmeisterBartelsMosler_2018}, an incrementally defined energy is introduced whose minimizers define all involved variables jointly. These minimizers also encompass the respective balance laws. By adding gradient terms of the plastic slips to the Helmholtz energy, non-local gradient-enhanced models are naturally included in the aforementioned variational setting, cf. \cite{FohrmeisterBartelsMosler_2018}. This allows to extend algorithms originally designed for classic local models also to gradient-enhanced theories. This train of thoughts is also taken here. The main contributions presented in this paper are:
\begin{itemize}
	\item Numerically robust algorithmic formulations suitable for local rate-independent crystal plasticity theory based on incremental energy minimization / variational constitutive updates. 
	\begin{itemize}
		\item The first algorithm falls into the range of nonlinear complementarity problem (NCP) functions. In order to improve the numerical performance, a new scaling of the NCP-function is proposed. Furthermore, it is shown that a regularized Fischer-Burmeister NCP function also eliminates the aforementioned (local) non-uniqueness of classic rate-indendendent crystal plasticity theory by implicitly modifying the underlying model.
		\item The new scaling of the regularized Fischer-Burmeister NCP function is also integrated into a classic, non-variational return-mapping scheme resulting in a robust algorithm.
		\item If the Fischer-Burmeister NCP function is replaced by the Min-NCP function, this algorithm is shown to be identical to a numerical formulation based on an active-set strategy.
		\item The second algorithm falls into the range of the augmented Lagrangian framework. The advantages and drawbacks of different constraints are discussed and the numerically most beneficial one is identified. 
	\end{itemize}
	\item Unifying micromorphic extension of the local crystal plasticity models in order to capture non-local gradient effect. This extension is again based on incremental energy minimization and the resulting constrained optimization problems are solved by means of the algorithms already used for the local models.
\end{itemize}

The paper is organized as follows: Section~\ref{sec:crystal-plasticity} covers the fundamentals of crystal plasticity theory. It includes the kinematics, constitutive assumptions as well as a variational reformulation of this theory. While Subsection~\ref{subsec:local-crystal-plasticity} deals with the standard local theory, the incorporation of gradients effects through a micromorphic approach is discussed in Subsection~\ref{subsec:gradEnhanced}. Throughout Section~\ref{sec:crystal-plasticity}, a time-continuous setting is adopted. The time-discrete approximation of crystal plasticity is presented in Section~\ref{sec:incremental_energy_minimisation} --- again in a variational setting based on incremental energy minimization. Two numerically robust solution schemes for the resulting time-discrete non-linear constraint optimization problems are subsequently elaborated in Section~\ref{sec:Methods} --- the main part of the paper. Numerical examples provided in Section~\ref{sec:examples} highlight the properties of both novel algorithms.

\section{GRADIENT-ENHANCED CRYSTAL PLASTICITY THEORY IN A NUTSHELL}\label{sec:crystal-plasticity}

The fundamentals of gradient-enhanced crystal plasticity theory are concisely presented here. While Subsection~\ref{subsec:local-crystal-plasticity} deals with the classic local theory, the incorporation of the plastic slips' gradients is discussed in Subsection~\ref{subsec:gradEnhanced}. Subsection~\ref{subsec:local-crystal-plasticity} and Subsection~\ref{subsec:gradEnhanced} are structured in the same manner. The constitutive models are presented first (Subsections \ref{subsubsec:local-crystal-plasticity-constitutive} and \ref{subsubsec:gradEnhanced-constitutive}) and subsequently, a variational reformulation of the models is given (Subsections \ref{subsubsec:local-crystal-plasticity-variational} and \ref{subsubsec:gradEnhanced-variational}). Since all models are based on the same kinematic assumptions, these are presented in Subsection~\ref{subsubsec:local-crystal-plasticity-kinematics} first.

\subsection{Kinematics} \label{subsubsec:local-crystal-plasticity-kinematics}

In what follows, the deformation of reference configuration $ \mathcal{B}_0 $ to current configuration $ \mathcal{B}_t $ is measured by non-linear deformation map $\vec{\varphi}$. It maps point $\vec{X}\in\mathcal{B}_0$ to its counterpart $\vec{x}\in\mathcal{B}_t$. Within a local neighborhood, $\vec{\varphi}$ can be approximated by deformation gradient
\eql{
	\mat{F} := \GRAD\vec{\varphi}:=\frac{\partial\vec{x}}{\partial\vec{X}}
}
which is subjected to the local invertability constraint $\det\mat{F}>0$.

In line with standard multiplicative plasticity theory (see, e.g. \cite{Lee1969,Simo1998}), deformation gradient $\mat{F}$ is decomposed in an incompatible manner according to
\eql{
	\mat{F}= \F \el \cdot \F \pl \; , \; \mbox{with} \;  \det(\mat{F}\el) > 0 \; \mathrm{and} \; \det(\mat{F}\pl) > 0 .
	\label{eq-ffefp}
}
Here, $\mat{F}\pl$ is associated with plastic deformations, while the elastic distortion of the underlying atomic lattice is captured by $\mat{F}\el$. Decomposition~\eqref{eq-ffefp} allows to introduce the material frame indifferent elastic measure
\eql{
	\mat{C}\el := \transp{[\mat{F}\el]} \cdot \mat{F}\el
}
representing the elastic right Cauchy-Green tensor. While $\F \el$ (or $\mat{C}\el$ respectively) thus governs the elastic response of the considered material, plastic deformations are defined by means of a flow rule, i.e., an evolution equation. Denoting the normal vector of slip plane $(i)$ as $ \vec{N}\ssi $ and the respective slip direction as $ \vec{M}\ssi $, the aforementioned evolution equation can be written as
\eq{eq-kin-Lp}{
	\mat{L}\pl:=\dot{\mat{F}}\pl\cdot{[\mat{F}\pl]}^{-1}=\sum\limits_{i=1}^{\nsys}\dot{\gamma}\ssi\;\vec{M}\ssi\otimes\vec{N}\ssi \, .
}
Here, the superposed dot represents the material time derivative, $ {\gamma}\ssi $ is the plastic slip at slip plane $(i)$ and $\mat{L}\pl$ is the plastic part of the velocity gradient with respect to the intermediate configuration induced by multiplicative split~\eqref{eq-ffefp}. Accordingly, the flow rule is also defined with respect to this intermediate configuration. Finally, it is noted that this intermediate configuration is isoclinic in the sense that vectors $\vec{M}\ssi$ and $\vec{N}\ssi$ are identical to their counterparts in the reference configuration.

\subsection{Local crystal plasticity theory}
\label{subsec:local-crystal-plasticity}

Throughout this paper, only models with associative evolution equations are considered. They can be defined by means of only two potentials: Helmholtz energy $\Psi$ as well as yield function $\phi$. Alternatively and referring to variational settings such as those in \cite{Ortiz_Stainier:99,Ortiz_Repetto:99,Carstensen_Hackl_Mielke:02,Miehe:02,Mosler-Bruhns2010,Stainier2010,FohrmeisterBartelsMosler_2018}, these models can also be derived by postulating Helmholtz energy $\Psi$ and dissipation potential $\Dint$.

\subsubsection{Constitutive model} \label{subsubsec:local-crystal-plasticity-constitutive}

Within plasticity theory, the elastic properties are not affected by inelastic deformations. This corresponds to an additive decomposition of the Helmholtz energy. Accordingly, $\Psi$ is assumed of type
\eq{eq-helmholtz-local}{
	\Psi(\mat{F}\el,\vec{\alpha})=\Psi\el(\mat{C}\el)+\Psi\pl(\vec{\alpha}).
}
While $\Psi\el$ defines the elastic response of the material, work due to hardening is accounted for by $\Psi\pl$. The latter energy depends on strain-like internal variables $ \vec{\alpha}$. Based on Eq.~\eqref{eq-helmholtz-local}, the dissipation of the local plasticity model is given by
\eq{eq-Dissipation}{
	\Dint = \tens{P} :  \dot{\mat{F}} - \dot{\Psi} = \tens{P} :  \dot{\mat{F}} - \frac{\partial\Psi\el}{\partial\mat{F}} : \dot{\tens{F}} - \frac{\partial \Psi\pl}{\partial \vec{\alpha}} \cdot \dot{\vec{\alpha}}\ge 0\, ,
}
where $ \tens{P} $ denotes the first Piola-Kirchoff stress tensor. The scalar product occurring in the last term of Eq.~\eqref{eq-Dissipation} apparently depends on the properties of $\vec{\alpha}$. Application of the by now classic Coleman and Noll procedure (see \cite{ColemanNoll1963}) yields
\eq{eq-P}{
	\mat{P}=\frac{\partial\Psi\el}{\partial\mat{F}}=\frac{\partial\Psi\el}{\partial\mat{F}\el}:\frac{\partial\mat{F}\el}{\partial\mat{F}}=2\;\mat{F}\el\cdot\frac{\partial\Psi\el}{\partial\mat{C}\el}\cdot{\mat{F}\pl}^{-T} \, .
}
which, in combination with notations
\eq{eq-internal-drivingforces}{
	\vec{Q} = -\frac{\partial \Psi\pl}{\partial\vec{\alpha}} \, , \qquad \, \mat{\Sigma}=2\;\mat{C}\el\cdot\dfrac{\partial\Psi\el}{\partial\mat{C}\el}
}
leads to reduced dissipation inequality
\eq{eq-Dissipation-red}{
	\Dint^\mathrm{red} = \tens{\Sigma} :  \mat{L}\pl + \vec{Q}\cdot\dot{\vec{\alpha}}\ge 0.
}
In Eq.~\eqref{eq-internal-drivingforces} and \eqref{eq-Dissipation-red}, $\mat{\Sigma}$ are the Mandel stresses (with respect to the intermediate configuration) and $\vec{Q}$ are internal stress-like variables dual to their strain-like counterparts $\vec{\alpha}$.

The constitutive assumptions summarized before are not restricted to crystal plasticity theory, but hold for a more general broad class of plasticity models. Focusing on crystal plasticity theory next, internal variable $\vec{\alpha}$ is chosen as $\vec{\alpha}=\{\alpha^{(1)},\ldots,\alpha^{(\nsys)}\}$ where $\alpha^{(i)}$ is a strain-like internal variable capturing hardening at slip plane $(i)$. The stress-like internal variables dual to $\alpha^{(i)}$ are denoted as $Q^{(i)}:=-\partial_{\alpha^{(i)}}\Psi$. With these notations and by introducing Schmid stresses
\eql{
	\taui = \vec{M}\ssi \cdot \mat{\Sigma} \cdot \vec{N}\ssi
}
yield functions of type
\eq{eq-yield-local-20}{
	\phi\ssi(\taui,Q\ssi) = \left|\taui\right| - (Q_0\ssi+Q\ssi)\le 0
}
are defined. In Eq.~\eqref{eq-yield-local-20}, $Q_0\ssi $ is the initial yield limit of slip system $(i)$. The intersection of all constraints~\eqref{eq-yield-local-20} results in the space of admissible stresses
\eq{eq-admissible-stress-space}{
	\mathbb{E}:=\left\{\left. (\mat{\Sigma},\vec{Q})\in\mathbb{R}^{9+\nsys}\; \right|\;\phi\ssi(\mat{\Sigma},\vec{Q})\le 0\;\forall i\in\{1,\ldots,\nsys\}\right\}.
}
$\Psi$ and $\mathbb{E}$ (respectively, $\phi\ssi(\taui,Q\ssi)$) define the constitutive model completely, if associative evolution equations are chosen. More specifically, application of the postulate of maximum dissipation leads to the well-known evolution equations 
\eq{eq-evolutions}{
	\dot{\alpha}\ssi=\lambda\ssi\;\partial_{Q\ssi}\phi\ssi=-\lambda\ssi \, \text{,} \qquad \mat{L}\pl=\sum_{i=1}^{\nsys} \lambda\ssi\;\partial_{\kmat{\Sigma}}\phi\ssi=
	\sum_{i=1}^{\nsys} \lambda\ssi\;\sign\left( \schmid\ssi \right) \;\vec{M}\ssi\otimes\vec{N}\ssi
}
depending on plastic multipliers $\lambda^{(i)}$. According to Eq.~\eqref{eq-evolutions}, these multipliers are identical to the negative rates of $\alpha^{(i)}$ for the considered constitutive model. Within the variational framework of the postulate of maximum dissipation, loading and unloading follow from the classic Karush-Kuhn-Tucker conditions
\begin{align}
	\phi\ssi & \leq 0 \, , &&
	\lambda\ssi \geq 0\, , &&
	\phi\ssi\, \lambda\ssi = 0 \, .
	\label{eq-maxD_KKT}
\end{align}
Although the structure of crystal plasticity theory is relatively simple, the computation of the set of active slip systems 
\eql{
	\aset := \left\lbrace i \in 1,\dots,\nsys \, | \lambda^{(i)}> 0 \right\rbrace \, .
}
and the computation of $ \lambda\ssi$ are very challenging for rate-independent models such as those discussed here.

\subsubsection{Variational structure of the model ---  time-continuous setting}
\label{subsubsec:local-crystal-plasticity-variational}

The variational structure of the model summmarized before (postulate of maximum dissipation) can be even further highlighted --- as for instance shown in \cite{Ortiz_Stainier:99,Ortiz_Repetto:99,Carstensen_Hackl_Mielke:02,Miehe:02,Berdichevsky_2006,Mosler-Bruhns2010,Stainier2010,FohrmeisterBartelsMosler_2018}. For that purpose, rate potential 
\eq{eq-potrate}{
	\potrate \coloneqq \integral{\bodyref}{}{ \left[
		\dot{\Psi} + \Dint
		\right] }{V} - \mathcal{P}_{\disp}
}
is defined. It depends on the stress power (since $ \mat{P}:\dot{\mat{F}} = \dot{\Psi} + \Dint $) and on power $ \mathcal{P}_{\disp} $ due to externally applied forces (i.e.\ volume or traction forces). It bears emphasis that functional~\eqref{eq-potrate} is considered to be a functional in rates $\dot{\vec{\varphi}}$ and $\lambda^{(i)}$. To be more explicit, the state itself ($\vec{\varphi}$, $\mat{F}\pl$ and $\alpha^{(i)}$) is not varied in Eq.~\eqref{eq-potrate}, i.e., $\potrate=\potrate(\dot{\vec{\varphi}},\lambda^{(1)},\ldots,\lambda^{(\nsys)})$. The key role of potential~\eqref{eq-potrate} becomes evident, if the Euler-Lagrange equations of variational principle
\eq{eq-minpotrate}{
	\stat_{\dot{\disp},\, \vec{\lambda}} \left(\potrate \, | \,(\mat{\Sigma},\vec{Q})\in\mathbb{E} \right) \, .
}
are computed. These result in balance of linear momentum
\eq{eq-diff-potrate-disp}{
	\delta_{\dot{\disp}}{\potrate} = \integral{\bodyref}{}{\mat{P} : \delta \dot{\mat{F}} }{V} - \delta{\Pdisp} = 0 \qquad \forall \, \delta \dot{\disp} \, .
}
and
\eq{eq-diff-potrate-lambda}{
	\delta_{\lambda\ssi}\potrate = -\integral{\bodyref}{}{\phi\ssi \, \delta \lambda\ssi}{V}=0 \quad \forall \, \delta \lambda\ssi.
}
Here, $\delta_{(\bullet)}\potrate$ is the variation of $\potrate$ with respect to $(\bullet)$ --- not the variational derivative. Eq.~\eqref{eq-diff-potrate-lambda} covers elastic unloading as well as elastoplastic loading. Within the first case, the respective plastic multiplier is constant resulting in $\delta\lambda^{(i)}=0$, while the necessary condition for elastoplastic loading yields $\phi^{(i)}=0$. Consequently, $\phi^{(i)}\,\delta\lambda^{(i)}=0$ always holds, i.e., Eq.~\eqref{eq-diff-potrate-lambda}. In contrast to the postulate of maximum dissipation, variational principle~\eqref{eq-minpotrate}, however, does not only define the constitutive model, but also encompass the underlying balance equation, cf. Eq.~\eqref{eq-diff-potrate-disp}. 

It bears emphasis that evolution equation~\eqref{eq-evolutions}$_2$ was already inserted into Eq.~\eqref{eq-potrate} for the derivation of Eq.~\eqref{eq-diff-potrate-lambda}. However and as shown in \cite{Ortiz_Stainier:99,Carstensen_Hackl_Mielke:02}, one can also derive the complete set of evolution equations from variational principle $\text{stat}~\potrate$.

\subsection{Gradient Enhanced crystal plasticity theory -- Micromorphic approach}\label{subsec:gradEnhanced}

Next, the local crystal plasticity model presented before is extended in order to account for gradient effects. This is implemented by means of the so-called micromorphic approach, cf. \cite{Forest_2009,Rys2022b}.

\subsubsection{Constitutive model}\label{subsubsec:gradEnhanced-constitutive}

Gradient effects are often modeled by incorporating $\nabla\mat{F}\pl:=\partial\mat{F}\pl/\partial\mat{X}$ (or $\text{curl}\mat{F}\pl$) or $\nabla\alpha^{(i)}$ into Helmholtz energy. In what follows, $\nabla\alpha^{(i)}$ is chosen. However, instead of inserting $\nabla\alpha^{(i)}$ directly, the gradient model is approximated by means of the micromorphic approach, (see cf. \cite{Forest_2009,FohrmeisterBartelsMosler_2018}). For this reason, $\alpha^{(i)}$ is replaced by slack variable $s^{(i)}$. Deviations between $\alpha^{(i)}$ and $s^{(i)}$ are penalized by term
\eq{penalty-term-1}{
	\Psi^{\text{pen}} (\vec{\alpha},\vec{s})=\frac{1}{2}\, c_1\;\sum\limits_{i=1}^{n_\text{sys}} \left(\alpha^{(i)}-s^{(i)}\right)^2.
}
Clearly, other penalty functions are admissible as well. Here, $c_1$ plays the role of a penalty parameter and thus, it has to be chosen sufficiently large. With Eq.~\eqref{penalty-term-1}, gradient-enhanced Helmholtz energy
\eq{original-gradient-energy-1}{
	\Psi(\mat{F}\el,\vec{\alpha},\gradd\vec{\alpha})=
	\Psi\el(\mat{C}\el)
	+ \Psi\pl(\vec{\alpha})
	+ \Psi^{\text{p,nonl}}(\nabla\vec{\alpha}) \, .
}
is approximated by micromorphic model
\eq{original-gradient-energy-1-approx}{
	\Psi(\mat{F}\el,\vec{\alpha},\vec{s},\gradd\vec{s})=
	\Psi\el(\mat{C}\el)
	+ \Psi\pl(\vec{\alpha})
	+\Psi^{\text{pen}} (\vec{\alpha},\vec{s})
	+ \Psi^{\text{p,nonl}}(\nabla\vec{s}) \, .
}
where $ \Psi^{\text{p,nonl}}$ captures gradient-hardening.

The main advantage of micromorphic approximation~\eqref{original-gradient-energy-1-approx} is its relatively simple numerical implementation. To be more precise, although additional fields $s^{(i)}$ have to be discretized, e.g., by means of finite elements, the inequalities characteristic of plasticity theory (such as $\phi^{(i)}\le 0$) can still be checked point-wise and do not have to be considered at the global structural level as it is required for original model~\eqref{original-gradient-energy-1}. Furthermore, the extension of the local model to its gradient-enhanced version is straightforward for approximation~\eqref{original-gradient-energy-1-approx}.

\subsubsection{Variational structure of the model ---  time-continuous setting}\label{subsubsec:gradEnhanced-variational}

By replacing Helmholtz energy~\eqref{eq-helmholtz-local} by its gradient-enhanced version~\eqref{original-gradient-energy-1-approx}, and having in mind that additional fields $s^{(i)}$ have been introduced, variational principle~\eqref{eq-minpotrate} reads now
\eq{eq-minpotrate-grad}{
	\stat_{\dot{\disp},\, \vec{\lambda}, \, \dot{\vec{s}}} \left(\potrate \, | \,(\mat{\Sigma},\vec{Q})\in\mathbb{E} \right) \, .
}
Due to the additive structure of Eq.~\eqref{original-gradient-energy-1-approx}, stationarity with respect to deformation rate $\dot{\vec{\varphi}}$ still results in balance of linear momentum, while a variation with respect to $\dot{\alpha}^{(i)}$ (or $\lambda^{(i)}$) leads to the slightly extended version of Eq.~\eqref{eq-diff-potrate-lambda}
\eq{eq-diff-potrate-lambda-ext}{
	\delta_{\lambda\ssi}\potrate = -\integral{\bodyref}{}{
		\left[\phi\ssi -c_1\,\left(\alpha^{(i)}-s^{(i)}\right)\right]
		\, \delta \lambda\ssi}{V}=0 \quad \forall \, \delta \lambda\ssi.
}
Finally, a variation with respect to $\dot{\vec{s}}$ yields
\eq{eq-diff-potrate-phip}{
	\delta_{\dot{s}^{(i)}}\potrate = -\integral{\bodyref}{}{\left[ \frac{\partial \Psi}{\partial s^{(i)}} - \DIV \frac{\partial \Psi}{\partial \nabla s^{(i)}} \right]\, \delta \dot{s}^{(i)}}{V} = 0 \qquad \forall \, \delta \dot{s}^{(i)}
}
where homogeneous Neumann boundary conditions have been assumed for $s^{(i)}$. Eq.~\eqref{eq-diff-potrate-phip} is often also denoted as balance of microforces, cf. \cite{Gurtin1996}. By inserting derivative $\partial \Psi/\partial s^{(i)}=-c_1(\alpha^{(i)}-s^{(i)})$ into Eq.~\eqref{eq-diff-potrate-phip} and by applying the localization theorem, one obtains the local counterpart of Eq.~\eqref{eq-diff-potrate-phip}
\eq{eq-diff-potrate-phip-loc}{
	-c_1(\alpha^{(i)}-s^{(i)}) - \DIV \frac{\partial \Psi}{\partial \nabla s^{(i)}} =0
}
In order to interpret the model better, Eq.~\eqref{eq-diff-potrate-phip-loc} and yield function~\eqref{eq-yield-local-20} are inserted into Eq.~\eqref{eq-diff-potrate-lambda-ext} and an elasto-plastic loading step is considered (i.e., $\delta\lambda^{(i)}\not =0$). In this case, application of the localization theorem results in
\eq{eq-yield-nonlocal-1}{
	\phi^{(i),\text{nonl}}:=
	\phi^{(i)}-\DIV \frac{\partial \Psi}{\partial \nabla s^{(i)}}=
	\left|\taui\right| -\Bigg(Q_0\ssi+\underbrace{Q\ssi+\DIV \frac{\partial \Psi}{\partial \nabla s^{(i)}}}_{\D =Q^{(i),\text{nonl}}}\Bigg)\le 0
}
which clearly highlights the gradient-enhancement of the yield function (last term in Eq.~\eqref{eq-yield-nonlocal-1}). Accordingly, the model shows an additional non-local hardening part in the yield function.

\section{Variational constitutive updates / incremental energy minimisation --- time-discrete setting}
\label{sec:incremental_energy_minimisation}

So far, a time-continuous setting has been adopted. Next, focus is on the numerical implementation in which the fields are approximated in space by means of a finite elements and in time by means of a suitable time integration. As a consequence, a (time) discrete setting is considered in this section.

\subsection{Discretization}
\label{subsec:incremental_energy_minimisation-discretization}

In order to obtain a numerical solution for time-continuous problems \eqref{eq-minpotrate} and \eqref{eq-minpotrate-grad}, a suitable discretization in time is required. Here, a fully implicit scheme is applied to time interval $[t_{n-1};t_n]$. Indices denoting objects of the current time step $ t_n $ are omitted for better readability, i.e., $t:=t_n$.

With the integrated plastic multipliers $\Delta\lambda\ssi := \int_{t_{n-1}}^{t}\lambda\ssi\,\text{d}t$, evolution equations \eqref{eq-evolutions} result in
\eq{eq-evolutions-discretised}{
	\alpha\ssi = -\Delta \lambda\ssi + \alpha\ssi_{n-1} \, , \qquad
	\gamma\ssi = \underbrace{\Delta \lambda\ssi \, \sign \left( \tau\ssi \right)}_{= \Delta\gamma\ssi} + \gamma\ssi_{n-1} \, , \qquad
	\Delta \Lp = \sum_{i \in \aset} \Delta\gamma^{(i)} \vec{M}\ssi\otimes\vec{N}\ssi \, .
}
The deviatoric structure of the flow rule ($\text{tr}\mat{L}\pl=0$) can be preserved in the time discrete setting by employing an implicit time discretization by means of the exponential map, i.e.,
\eq{eq-evolFp-Expmap}{
	\mat{F}\pl=\EXP \left( \Delta\Lp \right)\cdot
	\mat{F}\pl_{n-1} \, .
}
Alternatively, a classic backward-Euler integration without an additional projection step has also been implemented. With these time discretizations, time-continuous potential~\eqref{eq-potrate} (and its gradient-enhanced counterpart~\eqref{eq-minpotrate-grad}) leads to time-dependent incremental potential
\eq{eq-potinc}{
	\pot = \integral{\bodyref}{}{ \left[
		\Psi - \Psi_{n-1} + \Delta t\, \Dint
		\right] }{V} - \mathcal{E}_{\disp} \, ,
}
where $ \mathcal{E}_{\disp} $ denotes the work done by externally applied forces. 

Having discussed the time discretization, focus is on the spacial discretization next. In line with a standard isoparametric finite element scheme, both reference coordinates $ \vec{X} $ as well as placement field $ \disp $ and micromorphic field $ s^{(i)} $ are spanned by means of a $ C_0 $-continuous ansatz, i.e., finite element approximations 
\begin{equation}
	\vec{X}_h=\sum\limits_{k=1}^{n_\mathrm{e}} N_{(k)}\,\vec{X}_{(k)},\qquad
	\boldsymbol{\varphi}_h = \sum\limits_{k=1}^{n_\mathrm{e}} N_{(k)}\,\boldsymbol{\varphi}_{(k)},\qquad
	s^{(i)}_h = \sum\limits_{k=1}^{n_\mathrm{e}} N_{(k)}\,s^{(i)}_{(k)}
\end{equation}
are made. Here, $ n_\mathrm{e} $ is the number nodes per element, $N^{(k)}$ is the shape function associated with node $(k)$ and $\vec{X}_{(k)}$, $\disp_{(k)}$ and $s^{(i)}_{(k)}$ denote nodal coordinates of the reference configuration, the placement field and slack variable $s^{(i)}$, respectively. Clearly, other approximation schemes are also possible. For the sake of completeness, the variations and the gradients of $\boldsymbol{\varphi}_h$ and $s^{(i)}_h$ are also given. They result in
\begin{equation}\label{variations-fem-1}
	\nabla\boldsymbol{\varphi}_h = \sum\limits_{k=1}^{n_\mathrm{e}} \boldsymbol{\varphi}_{(k)}\otimes\nabla N_{(k)}=\mat{F}_h,\qquad
	\nabla\delta\boldsymbol{\varphi}_h = \sum\limits_{k=1}^{n_\mathrm{e}} \delta\boldsymbol{\varphi}_{(k)}\otimes\nabla N_{(k)}=\delta\mat{F}_h
\end{equation}
and
\begin{equation}\label{variations-fem-2}
	\nabla s^{(i)}_h = \sum\limits_{k=1}^{n_\mathrm{e}} s^{(i)}_{(k)}\otimes \nabla N_{(k)},\qquad
	\nabla \delta s^{(i)}_h = \sum\limits_{k=1}^{n_\mathrm{e}} \delta s^{(i)}_{(k)}\otimes \nabla N_{(k)}.
\end{equation}

\subsection{Discrete Minimisation Problem} \label{subsec:incremental_energy_minimisation-minima}

By inserting the spatial as well as the temporal discretizations into Eq.~\eqref{eq-potinc}, the incremental energy's contribution of finite element $e$ reads
\eq{eq-potinc-discr-1}{
	\mathcal{I}^e_{\text{inc}} 
	= \mathcal{I}^e_{\text{inc}}(\boldsymbol{\varphi}_h,s^{(1)}_h,\ldots,s^{(\nsys)}_h,\Delta\lambda^{(1)},\ldots,\Delta\lambda^{(\nsys)}).
}
The system's total energy can then be computed by a summation over all elements. Furthermore, continuity of fields $\vec{\varphi}_h$ and $s^{(i)}_h$ is enforced in standard manner by means of an assembly step. Clearly energy $\mathcal{I}^e_{\text{inc}}$ depends on the state at time $t_{n-1}$ as well as on state at time $t_n$. However, only the latter is unknown and thus the dependency on the known state at time $t_{n-1}$ is not highlighted explicitly. Energy~\eqref{eq-potinc-discr-1} depends on globally defined fields $\vec{\varphi}_h$ and $s^{(i)}_h$, while plastic multipliers $\Delta\lambda^{(i)}$ have to be defined only at the integrations points, since they do not enter the energy through gradients. For this reason, the plastic multipliers can be conveniently computed first from a local optimization problem. The respective optimization problem reads
\eq{plastic-multiplier-computation-1}{
	\Delta\lambda^{(i)}=\text{arg}\min\limits_{\Delta\lambda^{(i)}\ge 0} \mathcal{I}^e_{\text{inc}}.
}
This, in turn, defines reduced potential
\begin{equation}
	\pot^\mathrm{red}(\boldsymbol{\varphi}_h,s^{(1)}_h,\ldots,s^{(\nsys)}_h)= \min_{\dlambda\ssi\ge 0}  \pot \qquad\forall \, i \in 1,\dots,\nsys \label{eq-local-minimisation}.
\end{equation}
Consistency of condition~\eqref{plastic-multiplier-computation-1} (and thus, of potential~\eqref{eq-local-minimisation}) can be shown by computing the necessary extremum condition. A straightforward calculation yields
\eq{eq-ex-dIdlamda}{
	\dIdlambda = - \mat{\Sigma} : \left[ \partial_{\dlambda\ssi} \F\pl \cdot \inverse{ \left[ \F\pl \right]} \right] + Q_0 + Q\ssi
	+c_1\,(\alpha^{(i)}-s^{(i)})\ge 0.
}
Obviously, this derivative is not equivalent to non-local yield function~\eqref{eq-yield-nonlocal-1}, since $\partial_{\dlambda\ssi} \F\pl \cdot \inverse{[\F\pl]} \not =\vec{M}^{(i)}\otimes\vec{N}^{(i)}\,\text{sign}(\tau^{(i)})$ (see also the Appendix). However, consistency has to be checked for $\Delta t=t_n-t_{n-1}\to 0$. In this case, a linearization of exponential maps gives
\eq{drivative-exp}{
	\frac{\partial\mat{F}\pl}{\partial\Delta\lambda^{(i)}}=\Bigg[
	\underbrace{\frac{\partial\text{EXP}(\Delta\mat{L}\pl)}{\partial\Delta\mat{L}\pl}}_{\D =\mathbb{I}\quad (\Delta t\to 0)}:\left(\vec{M}^{(i)}\otimes\vec{N}^{(i)}\right)\,\sign(\tau^{(i)})\Bigg]\cdot\mat{F}\pl_{n-1}
	\overset{ \Delta t \rightarrow 0}{\longrightarrow}
	\,\sign(\tau^{(i)})\,\left(\vec{M}^{(i)}\otimes\vec{N}^{(i)}\right)\cdot\mat{F}\pl
}
and consequently, Eq.~\eqref{eq-ex-dIdlamda} indeed converges to
\eq{qe-ex-dIdlambda-conv}{
	\dIdlambda \overset{ \Delta t \rightarrow 0}{\longrightarrow}
	- |\mat{\Sigma} : \left(\vec{M}^{(i)}\otimes\vec{N}^{(i)}\right)| + Q_0 + Q\ssi
	+c_1\,(\alpha^{(i)}-s^{(i)})=-\phi^{(i),\text{nonl}}\ge 0.
}
Accordingly, a minimization of energy $\mathcal{I}^e_{\text{inc}}$ with respect to the plastic multipliers enforces the stresses to be admissible, i.e., constraint $\phi^{(i),\text{nonl}}\le 0$ associated with yield function $(i)$ is automatically fulfilled. However, It bears emphasis that $\partial\mathcal{I}_{\text{inc}}/\partial\Delta\lambda^{(i)}\not =-\phi^{(i),\text{nonl}}$ for a finite time step such as that considered in numerical simulations. Thus, the presented variational algorithm is not equivalent to a standard return mapping algorithm in which $\phi^{(i),\text{nonl}}\le 0$ is always explicitly enforced. Nevertheless, the variational update indeed approximates the yield function consistently, see also \cite{Mosler-Bleier2012} and all schemes are first-order accurate.

Finally, the remaining stationary conditions of potential~\eqref{eq-potinc-discr-1} are derived. A variation of Eq.~\eqref{eq-potinc-discr-1} with respect to placement $\vec{\varphi}_h$ yields
\eq{discrete-balanace-law-1}{
	\delta_{\kmat{\varphi}_h}\mathcal{I}_{\text{inc}}=\int\limits_{\Omega}\underbrace{\frac{\partial\Psi}{\partial\mat{F}_h}}_{\D =\mat{P}_h}:\delta\mat{F}_h\,\text{d}V
	-\int\limits_{\Omega}\frac{\partial\mathcal{E}}{\partial\vec{\varphi}_h}:\delta\vec{\varphi}_h\,\text{d}V=0\qquad\forall\delta\vec{\varphi}_h
}
which is the weak for of equilibrium (i.e., balance of linear momentum, cf. Eq.~\eqref{eq-diff-potrate-disp}), while a variation with respect to slack variable $s_h^{(i)}$ results in
\eq{discrete-balanace-law-2}{
	\delta_{s_h^{(i)}}\mathcal{I}_{\text{inc}}=\int\limits_{\Omega}\frac{\partial\Psi}{\partial s_h^{(i)}}:\delta s_h^{(i)}\,\text{d}V
	+\int\limits_{\Omega}\frac{\partial\Psi}{\partial\nabla s^{(i)}_h}:\delta\nabla s_h^{(i)}\,\text{d}V=0\qquad\forall\delta s_h^{(i)}.
}
Eq.~\eqref{discrete-balanace-law-2} is the weak form of balance of microforces, cf. Eq.~\eqref{eq-diff-potrate-phip}. In summary, variational principle $\min\mathcal{I}_{\text{inc}}$ enforces canonically: (1) balance of linear momentum, (2) balance of micro forces and (3) admissibility of the stresses, i.e., the yield functions.

\section{Incremental energy minimization as constrained optimization problem} \label{sec:Methods}

According to the previous section, the discretized local as well as the gradient-enhanced plasticity model are characterized by a minimization problem of type
\eq{disc-minimization-unique-1}{
	\min\limits_{\mathbb{X}_h}\min\limits_{\Delta\lambda^{(1)},\cdots,\Delta\lambda^{(\nsys)}}\mathcal{I}_{\text{inc}}.
}
While the local model is defined by set
\eq{disc-minimization-unique-2}{
	\mathbb{X}_h:=\left\{\kmat{\varphi}_h\right\},
}
the gradient-enhanced model is obtained by
\eq{disc-minimization-unique-3}{
	\mathbb{X}_h:=\left\{\kmat{\varphi}_h,s_h^{(1)},\ldots,s^{(\nsys)}_h\right\}.
}
The minimization problem is sufficiently smooth and unconstrained in $\mathbb{X}_h$ and thus, it can be conveniently solved by means of Newton's method. For that purpose, weak form~\eqref{discrete-balanace-law-1} is considered for the local model, while weak forms~\eqref{discrete-balanace-law-1} and \eqref{discrete-balanace-law-2} are solved monolithically by finite elements as far as the gradient-enhanced model is concerned.

According to the previous paragraph, a minimization of $\mathcal{I}_{\text{inc}}$ with respect to $\mathbb{X}_h$ is relatively straightforward. By way of contrast, although minimization problem
\eq{disc-minimization-unique-4}{
	\min\limits_{\Delta\lambda^{(1)},\cdots,\Delta\lambda^{(\nsys)}}\mathcal{I}_{\text{inc}}
}
is local (and consequently, it can be solved at the integration points), it is non-smooth and constrained and thus, numerically indeed challenging.

\subsection{Nonlinear Complementarity Problem (NCP) functions} \label{subsec:Methods-FischerBurmeister}

\subsubsection{Fundamentals}

The conventional, non-variational, setting of (local) crystal plasticity theory is characterized by the set of (in)equalities
\eq{NCP-1}{
	\phi^{(i)}\le 0,\qquad\Delta\lambda^{(i)}\ge 0,\qquad
	\phi^{(i)}\,\Delta\lambda^{(i)}=0.
}
By setting $a=-\phi^{(i)}$ and $b=\Delta\lambda^{(i)}$ this set falls into the range of Nonlinear Complementarity Problems (NCP). That was observed in \cite{Schmidt-Baldassari:03} first. For this reason, one can replace set~\eqref{NCP-1} by NCP function $G$. To be more precise,
\eq{NCP-1-2}{
	a\ge 0,\qquad b\ge 0,\qquad a\,b=0\qquad\Leftrightarrow\qquad
	G(a,b)=0
}
see, e.g., \cite{Geiger_Kanzow:02} for further information. Accordingly, the inequalities are replaced by one equality. In \cite{Schmidt-Baldassari:03}, a Fischer-Burmeister NCP function was chosen for $G$.  An implementation, together with some discussions concerning the numerical robustness was presented in \cite{FohrmeisterDiazMosler_2018a}.

Two examples for NCP functions are the aforementioned Fischer-Burmeister function
\eq{NCP-example-1}{
	G_\text{FB}:=\sqrt{a^2+b^2}-a-b
}
and the Min-NCP function
\eq{NCP-example-2}{
	G_\text{min}:=\min (a,b).
}
Clearly, both functions are continuous but not differentiable at the origin. Hence, if Newton's method is applied in order to solve Eq.~\eqref{NCP-example-1} or Eq.~\eqref{NCP-example-2}, one deals essentially with a non-smooth Newton iteration. Interestingly the resulting algorithm shares thus some similarities with an active set strategy. As a matter of fact, one can show that the Min-NCP function and a certain active set algorithm are equivalent. This is shown explicitly in Subsection~\ref{subsecsub:MinNCP}. 

The non-smoothness of NCP functions can be eliminated by a suitable regularization, cf. \cite{Geiger_Kanzow:02}. For this reason, the regularized 
Fischer-Burmeister function
\eq{NCP-example-1-2}{
	G_\text{FB}:=\sqrt{a^2+b^2+2\,\delta}-a-b
}
will be considered within the next paragraph. In Eq.~\eqref{NCP-example-1-2}, $\delta$ is a smoothing parameter which has to be chosen sufficiently small.

\subsubsection{Variational formulation for gradient-enhanced crystal plasticity theory}
\label{sec-variational-ncp-gradient}

\paragraph{Implementation based on Fischer-Burmeister function}

While in \cite{Schmidt-Baldassari:03,FohrmeisterDiazMosler_2018a} NCP functions were used in order to solve classic (non-variational) local plasticity theory, the gradient-enhanced variational framework elaborated in the previous sections is considered here. The respective optimization problem~\eqref{disc-minimization-unique-4} can be solved by means of Karush-Kuhn-Tucker conditions
\eq{disc-minimization-unique-5}{
	\frac{\partial\mathcal{I}_\text{inc}}{\partial\Delta\lambda^{(i)}}\ge 0,\quad
	\Delta\lambda^{(i)}\ge 0,\quad
	\frac{\partial\mathcal{I}_\text{inc}}{\partial\Delta\lambda^{(i)}}\,\Delta\lambda^{(i)}=0.
}
Hence, by insertimg $a=\partial\mathcal{I}_\text{inc}/\partial\Delta\lambda^{(i)}$ and $b=\Delta\lambda^{(i)}$ into Eq.~\eqref{NCP-example-1-2}, one obtains
\eq{eq-FischerBurmeister-smoothed}{
	\Gfb = \sqrt{ (\dlambda\ssi)^2 + \left(\dIdlambda\right)^2  + 2\,\delta} - \dlambda\ssi - \dIdlambda = 0 \, .
}
It bears emphasis that although the original model is based on energy minimization and thus, its Hessian is guaranteed to be symmetric, the Fischer-Burmeister function does not inherit this property. Accordingly, if Eqs.~\eqref{eq-FischerBurmeister-smoothed} are to be solved, e.g., by means of Newton's method, a solver for unsymmetrical matrices is required.

\paragraph{Implementation based on scaled Fischer-Burmeister function} \label{para:scaled Fischer-Burmeister}

NCP functions are purely mathematically motivated. This can be seen, for instance, by observing that the physical units do not fit. To be more precise, while $\dlambda\ssi$ is strain-like, derivative $\partial\mathcal{I}_\text{inc}/\partial\Delta\lambda^{(i)}$ is energy-like. In order to eliminate the dimensions of the structure, $\mathcal{I}_\text{inc}$ is replaced by its volume-specific counterpart
\eq{eq-potinc-loc-2}{
	i_{\text{inc}}:=\left[
	\Psi - \Psi_{n-1} + \Delta t\, \Dint\right]
	\quad\Rightarrow\quad
	\pot = \int\limits_\bodyref i_{\text{inc}}\,\text{d}V - \mathcal{E}_{\disp} \, ,
}
and Eq.~\eqref{eq-FischerBurmeister-smoothed} is replaced by
\eq{eq-FischerBurmeister-smoothed-21}{
	\Gfb = \sqrt{ (\dlambda\ssi)^2 + \left(\frac{\partial i_\text{inc}}{\partial\Delta\lambda^{(i)}}\right)^2  + 2\,\delta} - \dlambda\ssi - \frac{\partial i_\text{inc}}{\partial\Delta\lambda^{(i)}} = 0 \, .
}
Now, term $\partial i_\text{inc}/\partial\Delta\lambda^{(i)}$ is stress-like (to be more precise, a material stiffness like the shear modulus). In order to eliminate the aforementioned physical inconsistency, parameter $w$ is introduced and Eq.~\eqref{eq-FischerBurmeister-smoothed-21} becomes
\eq{eq-FischerBurmeister-smoothed-212}{
	\Gfb = \sqrt{ (w\,\dlambda\ssi)^2 + \left(\frac{\partial i_\text{inc}}{\partial\Delta\lambda^{(i)}}\right)^2  + 2\,\delta} - w\,\dlambda\ssi - \frac{\partial i_\text{inc}}{\partial\Delta\lambda^{(i)}} = 0 \, .
}
The interesting point about the inconsistency of the original Fischer-Burmeister function and modified function~\eqref{eq-FischerBurmeister-smoothed-212} is that it offers a scaling which affects the condition number of the resulting Newton matrix. Therefore, one could search for an optimal scaling parameter $w$. This idea is precisely elaborated in what follows.

\paragraph{(Local)-uniqueness of Fischer-Burmeister-based algorithms --- geometrically linearized, local model without hardening}

In order to analyze Fischer-Burmeister-based algorithms, a geometrically linearized setting without hardening is adopted (implying a local model without hardening), cf. \cite{FohrmeisterDiazMosler_2018a}. Within this setting, a straightforward calculation shows
\eq{equivalence-1}{
	\frac{\partial i_{\text{inc}}}{\partial\Delta\lambda^{(i)}}=-\phi^{(i)}\ge 0
}
and thus, the variational update is equivalent to the classic return-mapping algorithm. Accordingly, the optimal scaling parameter to be derived is also valid for the classic return-mapping algorithm combined with the Fischer-Burmeister function. Due to identity~\eqref{equivalence-1}, Eq.~\eqref{eq-FischerBurmeister-smoothed-212} can be replaced by
\eq{eq-FischerBurmeister-smoothed-21232}{
	\Gfb = \sqrt{ (w\,\dlambda\ssi)^2 + \left(\phi^{(i)}\right)^2  + 2\,\delta} - w\,\dlambda\ssi +\phi^{(i)} = 0
}
where the yield functions depending on Cauchy-stresses $\mat{\sigma}$ are of type (local model without hardening)
\eq{uniqueness-1}{
	\phi^{(i)}=\left|\mat{\sigma}:\left(\mat{M}^{(i)}\otimes\vec{N}^{(i)}\right)\right|-Q_0^{(i)}\le 0.
}
Furthermore, the multiplicative decomposition of the deformation gradient is replaced by an additive decomposition of the engineering strains, i.e., $\mat{\varepsilon}=\mat{\varepsilon}\el+\mat{\varepsilon}\pl$ and the stresses follows from Hooke's model $\mat{\sigma}=\mathbb{C}:\mat{\varepsilon}\el$. For linear isotropic elasticity, the elastic moduli take the form
$\mathbb{C}=2\mu\, \mathbb{I}+\Lambda\,\mat{1}\otimes\mat{1}$ where $\Lambda$ and $\mu$ are the Lam\'e constants. Combining these assumption with a backward-Euler time integration for the evolution equations results in
\eq{uniqueness-2}{
	\mat{\sigma}=\mathbb{C}:\left(\mat{\varepsilon}-\mat{\varepsilon}\pl_{n-1}-
	\sum\limits_i\Delta\lambda^{(i)}\,\sign\left[\mat{\sigma}:\left(\vec{M}^{(i)}\otimes\vec{N}^{(i)}\right)\right]\,\left(\vec{M}^{(i)}\otimes\vec{N}^{(i)}\right)^{\mathrm{sym}}\right) \, .
}
By inserting Eq.~\eqref{uniqueness-2} into Eq.~\eqref{uniqueness-1}, the plastic multipliers can be computed. More explicitly, introducing trial Schmid stresses
\eq{uniqueness-3}{
	\tau^{\mathrm{tr,}(i)}_{n}:=\left(\vec{M}^{(i)}\otimes\vec{N}^{(i)}\right):\mathbb{C}:\left(\mat{\varepsilon}_{n}-\mat{\varepsilon}\pl_{n-1}\right)
}
and inserting Eq.~\eqref{uniqueness-2} into Ineq.~\eqref{uniqueness-1} yields
\eq{uniqueness-4}{
	\phi_{n}=\left|\tau^{\mathrm{tr,}(i)}_{n}-\left(\vec{M}^{(i)}\otimes\vec{N}^{(i)}\right):\mathbb{C}:
	\left[\sum\limits_i\Delta\lambda^{(i)}\,\sign\left[\mat{\sigma}_{n}:\left(\vec{M}^{(i)}\otimes\vec{N}^{(i)}\right)\right]\,\left(\vec{M}^{(i)}\otimes\vec{N}^{(i)}\right)^{\mathrm{sym}}\right]\right|-Q_0^{(i)}\le 0 \, .
}
In order to avoid different cases, $\mat{\sigma}_n:(\vec{M}^{(i)}\otimes\vec{N}^{(i)})\ge 0$ is assumed for the sake of simplicity. Then, Ineq.~\eqref{uniqueness-4} can be re-written as
\eq{uniqueness-5}{
	\phi^{(i)}_n=\tau^{\mathrm{tr,}(i)}_{n}-\mu\,\sum\limits_{j}\mathbb{F}_{ij}\,\Delta\lambda^{(j)}-Q_0^{(i)}\le 0
}
with
\eq{uniqueness-7}{
	\mathbb{F}_{ij}=
	\left\{\left(\vec{M}^{(i)}\cdot\vec{M}^{(j)}\right)\,\left(\vec{N}^{(i)}\cdot\vec{N}^{(j)}\right)
	+\left(\vec{M}^{(i)}\cdot\vec{N}^{(j)}\right)\,\left(\vec{N}^{(i)}\cdot\vec{M}^{(j)}\right)\right\}\,.
}
The structure of $\mathbb{F}_{ij}$ is a direct consequence of the elastic moduli. As shown in \cite{FohrmeisterDiazMosler_2018a}, purely geometry-related matrix $\mathbb{F}_{ij}$ has a rank of five and thus, the problem is ill-posed in general (depending on the number of active slip systems), cf. \cite{Zuo2012}. 

In what follows, Fischer-Burmeister function~\eqref{eq-FischerBurmeister-smoothed-21232} is solved by means of Newton's method. Hence, the derivatives of $\Gfb$ with respect to $\Delta\lambda^{(i)}$ are required. Inserting~\eqref{uniqueness-5} into Eq.~\eqref{eq-FischerBurmeister-smoothed-21232} results in 
\eq{uniqueness-8}{
	\begin{array}{lll}
		\displaystyle \frac{\partial\Gfb}{\partial\Delta\lambda^{(j)}} & \displaystyle =& \displaystyle \frac{1}{ \sqrt{ (w\,\dlambda\ssi)^2 + \left(\phi^{(i)}\right)^2  + 2\,\delta}}\,
		\left[w^2\,\Delta\lambda^{(i)}\,\delta_{ij}-\mu\,\phi^{(i)}\,\mathbb{F}_{ij}\right]-w\,\delta_{ij}
		-\mu\,\mathbb{F}_{ij}\\[0.2cm]
		& \displaystyle =& \displaystyle
		\underbrace{\left[ \frac{w^2\,\dlambda\ssi}{ \sqrt{ (w\,\dlambda\ssi)^2 + \left(\phi^{(i)}\right)^2  + 2\,\delta}}-w\right]}_{\text{\textcircled{1}}}\,\delta_{ij}
		-\underbrace{\left[ \frac{\phi^{(i)}}{ \sqrt{ (w\,\dlambda\ssi)^2 + \left(\phi^{(i)}\right)^2  + 2\,\delta}}+1\right]}_{\text{\textcircled{2}}}\,
		\mu\,\mathbb{F}_{ij}
	\end{array}
}
where derivative $\partial\phi^{(i)}/\partial\Delta\lambda^{(j)}=-\mu\,\mathbb{F}_{ij}$ and Kronecker symbol $\delta_{ij}$ have been used. Eq.~\eqref{uniqueness-8} already identifies the different terms related to (local) well-posedness and ill-posedness of the problem. Since the matrix representation of $\mathbb{F}_{ij}$ is singular, regularity of $\partial\Gfb/\partial\Delta\lambda^{(j)}$ is controlled by factor \textcircled{1}, which scales the identity matrix ($\delta_{ij}$). 

One often starts with predictor $\Delta\lambda^{(i)}=0$ within the first iteration. In this case, \textcircled{1}$=-w$ and assuming a sufficiently small regularization parameter $\left(\phi^{(i)}\right)^2  + 2\,\delta\approx \left(\phi^{(i)}\right)^2$ and thus, \textcircled{2}$\approx \text{sign}(\phi^{(i)})+1$. As a consequence, derivatives~\eqref{uniqueness-8} take the form
\eq{uniqueness-9}{
	\frac{\partial\Gfb}{\partial\Delta\lambda^{(j)}}=-\left[\text{sign}(\phi^{(i)}_n)+1\right]
	\mu\,\mathbb{F}_{ij}-w\,\delta_{ij},\qquad\text{ if }\Delta\lambda^{(i)}=0.
}
According to Eq.~\eqref{uniqueness-9}, parameter $w$ indeed allows to control the regularity of matrix $\partial\Gfb/\partial\Delta\lambda^{(j)}$ and it has to be chosen in the order of the shear modulus. By doing so, the first Newton iteration is well-posed. 

Having discussed the first iteration, the final converged iteration is analyzed next. It is characterized by $\Delta\lambda^{(i)}\ge 0$ and by Eq.~\eqref{eq-FischerBurmeister-smoothed-21232}. By rewriting Eq.~\eqref{eq-FischerBurmeister-smoothed-21232} as $\sqrt{ (w\,\dlambda\ssi)^2 + \left(\phi^{(i)}\right)^2  + 2\,\delta} = w\,\dlambda\ssi -\phi^{(i)}$ and inserting this equation into Eq.~\eqref{uniqueness-8} one obtains for the final converged iteration step
\eq{uniqueness-10}{
	\begin{array}{llll}
		\displaystyle\frac{\partial\Gfb}{\partial\Delta\lambda^{(j)}} & \displaystyle= & \displaystyle
		\text{\textcircled{1}}\,\delta_{ij}-\text{\textcircled{2}}\,\mu\,\mathbb{F}_{ij}\\
		\displaystyle\text{\textcircled{1}} & \displaystyle= & \displaystyle
		\displaystyle \frac{\phi^{(i)}\, w}{w\,\Delta\lambda^{(i)}-\phi^{(i)}}
		& \displaystyle\qquad\text{ if converged and }Eq.~\eqref{eq-FischerBurmeister-smoothed-21232}\\
		\displaystyle\text{\textcircled{2}} & \displaystyle= & \displaystyle
		\displaystyle \left[1-\frac{\phi^{(i)}}{w\,\Delta\lambda^{(i)}}\right]^{-1}
		& \displaystyle \qquad\text{ if converged and }Eq.~\eqref{eq-FischerBurmeister-smoothed-21232}
	\end{array}
}
If smoothing parameter $\delta$ is relatively small, \textcircled{2}$\approx 1$. However, factor \textcircled{2} is less interesting, since it scales  singular matrix $\mathbb{F}_{ij}$. By ways of contrast, regularity of tangent $\partial\Gfb/\partial\Delta\lambda^{(j)}$ is mainly governed by factor  \textcircled{1}. Clearly, the unregularized Fischer-Burmeister function ($\delta=0$) corresponds to $\phi^{(i)}=0$ (for the active yield functions) and hence, \textcircled{1}$=0$. As a consequence, the (unregularized) Fischer-Burmeister function yields
\eq{uniqueness-10-neu1-1}{
	\frac{\partial\Gfb}{\partial\Delta\lambda^{(j)}}=-\mu\,\mathbb{F}_{ij}=\frac{\partial\phi^{(i)}}{\partial\Delta\lambda^{(j)}}
	\qquad\text{ unregularized FB function ($\delta=0$) for the converged step}
}
and does not eliminate the (local) non-uniqueness of rate-independent crystal plasticity theory --- although that is sometimes claimed. More explicitly, the (unregularized) Fischer-Burmeister function shows the same tangent as the original underlying crystal plasticity model for a converged step.

In contrast to its unregularized version, the regularized Fischer-Burmeister function can regularize rate-independent crystal plasticity theory with respect to regularity of $\partial\Gfb/\partial\Delta\lambda^{(j)}$ even at the converged step. This can be seen by analyzing factor \textcircled{1} in Eq.~\eqref{uniqueness-10}. While the unregularized function is characterized by $\phi^{(i)}=0$, the regularized version is defined by Eq.~\eqref{eq-FischerBurmeister-smoothed-21232}. For this reason, $\phi^{(i)}\not =0$ at the converged step and likewise, \textcircled{1}$\not =0$. This can also be shown by squaring Eq.~\eqref{eq-FischerBurmeister-smoothed-21232} which, in turn, results in
\eq{uniqueness-10-neu1-1321}{
	\delta=-w\,\Delta\lambda^{(i)}\,\phi^{(i)}.
}
Eq.~\eqref{uniqueness-10-neu1-1321}, allows a direct mechanical interpretation of the regularized Fischer-Burmeister function. Since $\Delta\lambda^{(i)}\,\phi^{(i)}=0$ is the discrete persistency (or consistency) condition, Eq.~\eqref{uniqueness-10-neu1-1321} being equivalent to
\eq{uniqueness-10-neu1-13232321}{
	\Delta\lambda^{(i)}\,\phi^{(i)}=-\frac{\delta}{w}
}
is an approximation of this condition. Such an approximation can be found in many optimization algorithms -- for instance, in the interior point method, cf. \cite{SCHEUNEMANN20201,SCHEUNEMANN2021111149,Niehueser_Mosler:23}. By inserting Eq.~\eqref{uniqueness-10-neu1-1321} into the numerator $\phi^{(i)}\,w$ of factor \textcircled{1} in Eq.~\eqref{uniqueness-10}, one obtains 
\eq{uniqueness-10-neu1-1ewe321}{
	\phi^{(i)}\,w=-\frac{\delta}{\Delta\lambda^{(i)}}.
}
According to Eq.~\eqref{uniqueness-10-neu1-1ewe321}, $\delta$ has to be chosen as sufficiently large in order to obtain a regular matrix  $\partial\Gfb/\partial\Delta\lambda^{(j)}$, see factor \textcircled{1} in Eq.~\eqref{uniqueness-10}. However, $\delta$ must also not be chosen as too large, since otherwise the error between the regularized Fischer-Burmeister function and the original crystal plasticity model is not acceptable, see Eq.~\eqref{uniqueness-10-neu1-1ewe321}.

\paragraph{(Non)-uniqueness of Fischer-Burmeister-based algorithms --- \\ geometrically exact crystal plasticity theory with hardening}

For the more general case being gradient-enhanced, geometrically exact crystal plasticity theory with hardening, the derivative of the Fischer-Burmeister function reads
\eq{uniqueness-14321}{
	\frac{\partial\Gfb}{\partial\Delta\lambda^{(j)}}=\frac{1}{ \D \sqrt{ (w\,\dlambda\ssi)^2 + \left(\frac{\partial i_\text{inc}}{\partial\Delta\lambda^{(i)}}\right)^2  + 2\,\delta} }\,
	\left[w^2\,\Delta\lambda^{(i)}\,\delta_{ij}+\frac{\partial i_\text{inc}}{\partial\Delta\lambda^{(i)}}\,\frac{\partial^2 i_\text{inc}}{\partial\Delta\lambda^{(i)}\partial\Delta\lambda^{(j)}}\right]-w\,\delta_{ij}
	-\frac{\partial^2 i_\text{inc}}{\partial\Delta\lambda^{(i)}\partial\Delta\lambda^{(j)}}.
}
By using redefinitions $\mu\,\mathbb{F}_{ij}:=\partial^2 i_\text{inc}/\partial\Delta\lambda^{(i)}\partial\Delta\lambda^{(j)}$ and $\phi^{(i)}:=\partial i_\text{inc}/\partial\Delta\lambda^{(i)}$, Eq.~\eqref{uniqueness-14321} is equivalent to Eq.~\eqref{uniqueness-8} and thus, all properties derived for the previously considered geometrically linearized setting without hardening also hold in the more general setting discussed here. Clearly, matrix $\mathbb{F}_{ij}$ might show a slightly different structure in the case of geometrically exact theory. However, the elaborated regularization properties depend only on pre-factor \textcircled{1} in Eq.~\eqref{uniqueness-8} and this factor is equivalent for both settings. This equivalence leads to the following conclusion for the  geometrically exact crystal plasticity theory with hardening: (1) by choosing a proper scaling parameter $w$, the first Newton iteration is well-posed and (2) by choosing a proper regularization parameter $\delta$ also the final converged Newton iteration step is well-posed (matrix $\partial\Gfb/\partial\Delta\lambda^{(j)}$ is regular).

\paragraph{On the choice of the numerical parameters}
\label{choice-of-w}

As shown in the previous paragraphs, a sufficiently large parameter $w$ guarantees a well-posed Newton iteration for states sufficiently far away from the converged solution. Furthermore, $w$ couples (additively) strains and material stiffnesses like shear and bulk moduli. Since crystal plasticity is isochoric, the shear modulus is precisely a good and physically sound parameter for $w$, see also Eq.~\eqref{uniqueness-9}. Numerical experiments reported in Section~\ref{sec:examples} will confirm the good properties of choice $w=\mu$.

The second parameter involved in the regularized Fischer-Burmeister function is $\delta$. According to Eq.~\eqref{uniqueness-10} and \eqref{uniqueness-10-neu1-1}, if $\delta$ is sufficiently small, tangent $\partial\Gfb/\partial\Delta\lambda^{(j)}$ corresponding to the converged solution is equivalent to that of the underlying crystal plasticity model which has a rank deficiency and thus, the resulting set of equation becomes ill-posed. For this reason, $\delta$ has to be chosen as sufficiently large. However, $\delta$ must also not be chosen as too large, since otherwise the error between the regularized Fischer-Burmeister function and the original crystal plasticity model is not acceptable. Within the computations, regularization parameter $\delta$ has been set to $ \delta = \SI{1e-10}{} $ or $\sqrt{\delta}=10^{-5}$. Obviously, this cannot be a general recommendation, since the choice of $\delta$ depends on parameter $w$ as well as on the unit system defining the stresses, cf. Eq.~\eqref{uniqueness-10-neu1-1ewe321}. As shown in the numerical experiments in Section~\ref{sec:examples}, the aforementioned choices lead to matrices with sufficiently small condition numbers. In particular, the matrices are regular.

\subsubsection{Minimum Nonlinear Complementarity Problem interpreted vs. classic active set strategy} \label{subsecsub:MinNCP}

In addition to the Fischer-Burmeister function, an also frequently applied NCP function is the so-called Min-NCP function, cf. \cite{Geiger_Kanzow:02}. Considering the Karush-Kuhn-Tucker conditions
\eq{eq-minNCP-1-321}{
	a^{(i)}:=\frac{\partial\mathcal{I}_\text{inc}}{\partial\Delta\lambda^{(i)}}\ge 0,\quad
	b^{(i)}:=\Delta\lambda^{(i)}\ge 0,\quad
	\frac{\partial\mathcal{I}_\text{inc}}{\partial\Delta\lambda^{(i)}}\,\Delta\lambda^{(i)}=0.
}
the Min-NCP function reads
\eq{eq-minNCP-1}{
	\Gmin \coloneqq \min \left(a^{(i)},b^{(i)}\right)=0.
}
If these equations are solved by means of Newton's method, the linearization of $\Gmin$ is required. A straightforward calculation yields
\eq{eq-diff-minNCP-1}{
	\partial_{\dlambda^{(j)}} \Gmin=:\mathbb{H}^{(i,j)} = 
	\begin{cases} 
		\frac{\partial a^{(i)}}{\partial\Delta\lambda^{(j)}}\, & , \, a^{(i)}\le b^{(i)} \\
		\kronecker{i}{j} &, \, \text{else}.
	\end{cases}
}
By setting $a^{(i)}:=\partial\mathcal{I}_\text{inc}/\partial\Delta\lambda^{(i)}$ the variational constitutive update is obtained, while choice $a^{(i)}:=-\phi^{(i)}$ corresponds to a classic return-mapping scheme.

Without loss of generality, the integrated plastic multipliers are ordered such that $\Delta\lambda^{(i)}>0$ $\forall i\in\{1,\ldots,n_{\text{act}}\}$ and $\Delta\lambda^{(i)}=0$ $\forall i>n_{\text{act}}$. With this ordering, the update of $\Delta\lambda^{(i)}$ within Newton iteration $k$ is given by
\eq{eq-diff-minNCP-1322}{
	\begin{array}{ll}
		\D\Delta\lambda^{(j)}_{(k)}=\Delta\lambda^{(j)}_{(k-1)}-\left[\mathbb{H}^{(j,i)}\right]^{-1}\,\Gmin &\D\qquad\forall j\in\{1,\ldots,n_{\text{act}}\}\\
		\D\Delta\lambda^{(j)}_{(k)}=0&\D\qquad\forall i>n_{\text{act}}.
	\end{array}
}
Consequently, only the active slip systems evolve. For the non-variational algorithm $a^{(i)}:=-\phi^{(i)}$, the active slip systems evolve in the standard manner as predicted by Newton's method. In summary, the Min-NCP function is hence equivalent to a classic active set strategy in which the set of active slip systems is updated in every Newton iteration. As a result, it corresponds to non-smooth Newton methods. If other active set updates are considered such as updating after a converged Newton iteration, the aforementioned equivalence is lost.

\subsection{Augmented Lagrangian Approach --- Multiplier-Penalty-Methods} \label{subsec:Methods-AugLagrange}

Another powerful algorithm for solving constrained minimization problems is based on the augmented Lagrangian approach also known as Multiplier-Penalty-Method, cf. \cite{Geiger_Kanzow:02}.

\subsubsection{Fundamentals}

Focusing on (local) crystal plasticity theory, \cite{Schmidt-Baldassari:03} seems to be the first work in which an algorithm based on the augmented Lagrangian method was proposed. The minimization problem considered in \cite{Schmidt-Baldassari:03}  is the postulate of maximum dissipation, i.e.,
\eq{eq-lagrangian-form-1}{
	\max\Dint=\max\left[\mat{\Sigma}:\mat{L}\pl+\sum\limits_{i=1}^{n_{\text{sys}}}Q^{(i)}\,\dot{\alpha}^{(i)}\right]
	\quad\text{subject to}\quad\phi^{(i)}\le 0.
}
If constraints $\phi^{(i)}\le 0$ are implemented by means of the augmented lagrangian approach, textbooks such as \cite{Geiger_Kanzow:02} yield functional
\eq{eq-lagrangian-form-234}{
	\mathcal{L}_{a}=\mathcal{L}_{a}(\mat{\Sigma},Q^{(1)},\ldots,Q^{n_{\text{sys}}},\lambda^{(1)},\ldots,\lambda^{(n_{\text{sys}})})= \Dint  + \frac{1}{2\,\pvisco} \sum_{i=1}^{\nsys} \left[ {\max}^2 \left( 0,\lambda\ssi + \pvisco\, \phi\ssi \right) - \left[ \lambda\ssi \right]^2 \right] \, .
}
According to \cite{Schmidt-Baldassari:03}, $\alpha$ -- not to be confused with the strain-like internal variables $\alpha^{(i)}$ - is a penalty parameter and $1/\alpha$ can be interpreted as a (plastic) viscosity. However and in line with the augmented Lagrangian approach, if $\alpha$ is sufficiently large, it does not affect the solution. The stationary conditions associated with Eq.~\eqref{eq-lagrangian-form-234} are
\eq{eq-lagrangian-form-stat-1}{
	\begin{array}{ll}
		\D\partial_{\kmat{\Sigma}}\mathcal{L}_{a}=0 &\qquad \mat{L}\pl=\sum\limits_i\max\left[0,\lambda^{(i)}+\alpha\,\phi^{(i)}\right]\,\text{sign}(\tau^{(i)})\,\vec{M}^{(i)}\otimes\vec{N}^{(i)}\\
		\D\partial_{Q^{(i)}}\mathcal{L}_{a}=0 &\qquad \dot{\alpha}^{(i)}=-\max\left[0,\lambda^{(i)}+\alpha\,\phi^{(i)}\right]\\
		\D\partial_{\kmat{\Sigma}}\mathcal{L}_{a}=0 &\qquad \lambda^{(i)}=\max\left[0,\lambda^{(i)}+\alpha\,\phi^{(i)}\right].	
	\end{array}
}
In line with a standard implementation based on a classic return-mapping algorithm, $\lambda^{(i)}$ are thus the only unknowns. By inserting the integrated flow rule and internal variables $\alpha^{(i)}$ into Eq.~\eqref{eq-lagrangian-form-stat-1}$_3$ and by applying a backward Euler integration to Eq.~\eqref{eq-lagrangian-form-stat-1}$_3$ one obtains the update of $\Delta\lambda^{(i)}$ as (see \cite{Schmidt-Baldassari:03})
\eq{eq-auglagrange-form-153}{
	\Gal \coloneqq \dlambda\ssi_n - \max \left( 0, \, \dlambda\ssi_{n-1} + \dt \, \pvisco \, \phi\ssi(\dlambda\ssi_{n}) \right) \overset{!}{=} 0 \, .
}
Eq.~\eqref{eq-auglagrange-form-153} naturally explains the regularization property of the algorithm: The derivative of the first term $\partial\dlambda\ssi_n/\partial\dlambda_n^{(j)}=\delta_{ij}$ changes the spectrum of matrix $\partial\phi^{(i)}/\partial\dlambda^{(j)}_n$ and the linearized system of equations becomes well-posed (for a sufficiently large penalty parameter $\alpha$).

\subsubsection{Variational formulation for gradient-enhanced crystal plasticity theory}

Next, the augmented lagrangian approach is to be adopted for the gradient-enhanced crystal plasticity theory presented before. In this case, the variational principle is not the postulate of maximum dissipation, but
\eq{disc-minimization-unique-1-32}{
	\min\limits_{\mathbb{X}_h}\min\limits_{\Delta\lambda^{(1)},\cdots,\Delta\lambda^{(\nsys)}}\mathcal{I}_{\text{inc}}\qquad\text{with}\qquad
	\mathbb{X}_h:=\left\{\kmat{\varphi}_h,s_h^{(1)},\ldots,s^{(\nsys)}_h\right\}.
}
For this reason, a naive application of the augmented Lagrangian approach discussed in the previous paragraph reads
\eq{eq-lagrangian-form-2}{
	\mathcal{L}_a^I \coloneqq \Iinc  + \frac{1}{2\,\pvisco} \sum_{i=1}^{\nsys} \left[ {\max}^2 \left( 0,\beta\ssi + \pvisco\, \phi^{(i),\text{nonl}} \right) - \left[ \beta\ssi \right]^2 \right] \, .
}
where condition $\phi^{(i),\text{nonl}}\le 0$ is enforced by penalty parameter $1/\alpha$ and Lagrange multiplier $\beta^{(i)}$. Alternatively, one could enforce $\partial\Iinc/\partial\Delta\lambda^{(i)}\ge 0$ representing a variational approximation of $\phi^{(i),\text{nonl}}\le 0$. This would lead to
\eq{eq-lagrangian-form-3}{
	\mathcal{L}_a^{II} \coloneqq \Iinc  + \frac{1}{2\,\pvisco} \sum_{i=1}^{\nsys} \left[ {\max}^2 \left( 0,\beta\ssi - \pvisco\,
	\partial\Iinc/\partial\Delta\lambda^{(i)}\right) - \left[ \beta\ssi \right]^2 \right] \, .
}
However, both such augmented Lagrangians would not work, since condition $\partial\Iinc/\partial\Delta\lambda^{(i)}\approx -\phi^{(i),\text{nonl}}\ge 0$ is already included in unconstrained minimization problem~\eqref{disc-minimization-unique-1-32}. For this reason, $\phi^{(i),\text{nonl}}$ is simply replaced in Eq.~\eqref{eq-auglagrange-form-153} by $-\partial\Iinc/\partial\Delta\lambda^{(i)}$. As a result, the augmented Lagrangian of the variational crystal plasticity model is characterized by update
\eq{eq-auglagrange-form-1}{
	\Gal \coloneqq \dlambda\ssi_n - \max \left( 0, \, \dlambda\ssi_{n-1} - \dt \, \pvisco \,  \left.\partial\Iinc/\partial\Delta\lambda^{(i)}\right|_{t_n} \right) \overset{!}{=} 0 \, .
}
Within the implementation used for the numerical examples \eqref{eq-auglagrange-form-1} is solved by a Newton-type iteration scheme showing quadratic convergence. Subsequently, parameter $\alpha$ is updated until the solution does not change anymore within a prescribed tolerance.

\section{Numerical Examples}\label{sec:examples}

\begin{table}
	\begin{tabular*}{\textwidth}{c | c | c @{\hspace{10ex}} c | c | c}
		\hline
		$ i $ & $ \vec{M}\ssi $ & $ \vec{N}\ssi $ &         $ i $ & $ \vec{M}\ssi $ & $ \vec{N}\ssi $ \\
		1 & $ \left[-1,1,0\right]^T / \sqrt{2} $ & $ \left[1,1,1\right]^T / \sqrt{3} $ &
		13 & $ \left[1,-1,0\right]^T / \sqrt{2} $ & $ \left[1,1,1\right]^T / \sqrt{3} $ \\
		2 & $ \left[1,0,-1\right]^T / \sqrt{2} $ & $ \left[1,1,1\right]^T / \sqrt{3} $ &
		14 & $ \left[-1,0,1\right]^T / \sqrt{2} $ & $ \left[1,1,1\right]^T / \sqrt{3} $ \\
		3 & $ \left[0,-1,1\right]^T / \sqrt{2} $ & $ \left[1,1,1\right]^T / \sqrt{3} $ &
		15 & $ \left[0,-1,1\right]^T / \sqrt{2} $ & $ \left[1,1,1\right]^T / \sqrt{3} $ \\
		4 & $ \left[-1,-1,0\right]^T / \sqrt{2} $ & $ \left[1,-1,-1\right]^T / \sqrt{3} $ &
		16 & $ \left[1,1,0\right]^T / \sqrt{2} $ & $ \left[1,-1,-1\right]^T / \sqrt{3} $ \\
		5 & $ \left[1,0,1\right]^T / \sqrt{2} $ & $ \left[1,-1,-1\right]^T / \sqrt{3} $ &
		17 & $ \left[-1,0,-1\right]^T / \sqrt{2} $ & $ \left[1,-1,-1\right]^T / \sqrt{3} $ \\
		6 & $ \left[0,1,-1\right]^T / \sqrt{2} $ & $ \left[1,-1,-1\right]^T / \sqrt{3} $ &
		18 & $ \left[0,-1,1\right]^T / \sqrt{2} $ & $ \left[1,-1,-1\right]^T / \sqrt{3} $ \\
		7 & $ \left[1,1,0\right]^T / \sqrt{2} $ & $ \left[-1,1,-1\right]^T / \sqrt{3} $ &
		19 & $ \left[-1,-1,0\right]^T / \sqrt{2} $ & $ \left[-1,1,-1\right]^T / \sqrt{3} $ \\
		8 & $ \left[-1,0,1\right]^T / \sqrt{2} $ & $ \left[-1,1,-1\right]^T / \sqrt{3} $ &
		20 & $ \left[1,0,-1\right]^T / \sqrt{2} $ & $ \left[-1,1,-1\right]^T / \sqrt{3} $ \\
		9 & $ \left[0,-1,-1\right]^T / \sqrt{2} $ & $ \left[-1,1,-1\right]^T / \sqrt{3} $ &
		21 & $ \left[0,1,1\right]^T / \sqrt{2} $ & $ \left[-1,1,-1\right]^T / \sqrt{3} $ \\
		10 & $ \left[1,-1,0\right]^T / \sqrt{2} $ & $ \left[-1,-1,1\right]^T / \sqrt{3} $ &
		22 & $ \left[-1,1,0\right]^T / \sqrt{2} $ & $ \left[-1,-1,1\right]^T / \sqrt{3} $ \\
		11 & $ \left[-1,0,-1\right]^T / \sqrt{2} $ & $ \left[-1,-1,1\right]^T / \sqrt{3} $ &
		23 & $ \left[1,0,1\right]^T / \sqrt{2} $ & $ \left[-1,-1,1\right]^T / \sqrt{3} $ \\
		12 & $ \left[0,1,1\right]^T / \sqrt{2} $ & $ \left[-1,-1,1\right]^T / \sqrt{3} $ &
		24 & $ \left[0,-1,-1\right]^T / \sqrt{2} $ & $ \left[-1,-1,1\right]^T / \sqrt{3} $ \\
		\hline
	\end{tabular*}
	\caption{Normal vector $\vec{N}^{(i)}$ and slip direction $\vec{M}^{(i)}$ of slip system $i$ defining the structure of an FCC crystal. It bears emphasis that 24 slip systems are considered, i.e., for each slip system both directions $\pm\vec{M}^{(i)}$ are modeled by means of an individual yield function. The yield functions thus read $ \phi\ssi(\taui,Q\ssi) = \taui - (Q_0\ssi+Q\ssi)\le 0 $.}
	\label{tab-fcc-slipsystems}
\end{table}

The efficiency and the robustness of the two algorithms elaborated in the previous section are highlighted here by means of numerical examples. While a strain-driven local problem is investigated in Subsection~\ref{subsec:simpleshear} (single material point problem), a boundary value problem governed by gradient-enhanced crystal plasticity theory is numerically analyzed in Subsection~\ref{subsec:TensileTest}.

Within all computations, a Neo-Hooke-like, isotropic elastic energy of type
\eq{eq-neohooke}{
	\Psi\el = \frac{1}{2}\, \kappa \, [\log{\Je}]^2 + \frac{1}{2}\, \mu \, \left[ \trace \left(\bar{\mat C}\el \right) - 3 \right] \qquad \text{with} \qquad \bar{\mat C}\el = [\Je]^{-\frac{2}{3}} \, \mat{C}\el
}
is chosen. Hardening of the local plastic model is governed by an exponential functional defined by Helmholtz energy
\eq{eq-psipl}{
	\Psi\pl = \left[Q_\infty - Q_0\right] \, \left[A + H \, \exp \left( -\frac{A}{H} \right) - H \right]
}
where
\eql{
	A = \sum_{i=0}^{\nsys} |\alpha\ssi|
}
is the accumulated plastic slip of all slip systems. Furthermore, a face centered cubic crystal structure is considered. The respective slip directions and slip plane normal vectors are summarized in Tab.~\ref{tab-fcc-slipsystems}.

In the case of the micromorphic crystal plasticity model, hardening caused by plastic gradients is taken into account through energies
\eq{eq-psi-micromorph}{
	\Psi^{\text{pen}} = \frac{1}{2}\, c_1 \, \left[\alpha\ssi - s\ssi \right]^2 \qquad \text{and} \qquad
	\Psi^{\text{p,nonl}} = \frac{1}{2}\, c_2 \, \left[\GRAD s\ssi \cdot \vec{M}\ssi \right]^2 \, .
}
As already described in Section~\ref{subsec:gradEnhanced}, $\Psi^{\text{pen}}$ is a penalty function in order to approximate $s\ssi\approx\alpha\ssi$, while $\Psi^{\text{p,nonl}}$ corresponds to the gradient hardening model. According to Eq.~\eqref{eq-psi-micromorph}, hardening is assumed to be governed only by the component of gradient $ \varphi\ssi $ in slip direction. This is an approximation of the dislocation density related to edge dislocations, cf. \cite{Gurtin:06,Le2014}. Clearly, other hardening models can also be implemented, cf. \cite{Frank1951,NYE1953153,Steinmann2015-js}. 

\subsection{Simple Shear Loading Case}\label{subsec:simpleshear}
In line with \cite{Schmidt-Baldassari:03,Prger2020ACS}, an FCC crystal subjected to simple shear is numerically analyzed. The respective deformation gradient is given by $\mat{F}=\mat{1}+F_{12}\,\vec{e}_1\otimes\vec{e}_2$ where $ F_{12} \in[ 0,\, 4] $ is prescribed within the numerical simulations in equidistant load steps and $\vec{e}_i$ denote the cartesian basis vectors. In order to investigate the convergence properties of the algorithms, three different loading amplitudes with increments $ \Delta F_{12} = \num{2e-2} $, $ \num{2e-3} $ and $ \num{2e-4} $ are chosen. Since the prescribed deformation gradient corresponds to an affine deformation, gradient hardening is not active and thus, the computations can be restricted to the material point level. The parameters of the model are summarized in Table \ref{tab-matparam-shear}.
\begin{table}[ht!]
	\centering
	\begin{tabular}{|l r | l r | l r |}
		\hline
		$\kappa$        & \SI{49.98}{\giga\pascal} &
		$\mu$           & \SI{21.1}{\giga\pascal} &
		$Q_0$         & \SI{6e-2}{\giga\pascal} \\
		\hline
	\end{tabular}
	\caption{FCC crystal subjected to simple shear: model parameters. Perfect plasticity without hardening}
	\label{tab-matparam-shear}
\end{table}

Further following \cite{Schmidt-Baldassari:03,Prger2020ACS}, different initial orientations of the crystal are numerically analyzed. These orientations are prescribed in terms of an initial plastic deformation gradient $ \F\pl_\mathrm{init} = \mat{R}_0 \in SO(3) $ where rotation matrix
\eq{eq-init-rotation}{
	\mat{R}_0 = \begin{bmatrix}
		\cos c & \sin c & 0 \\
		-\sin c & \cos c & 0 \\
		0 & 0 & 1
	\end{bmatrix}
	\cdot
	\begin{bmatrix}
		1 & 0 & 0 \\
		0 & \cos b & \sin b \\
		0 & -\sin b & \cos b
	\end{bmatrix}
	\cdot
	\begin{bmatrix}
		\cos a & \sin a & 0\\
		-\sin a & \cos a & 0 \\
		0 & 0 & 1 \\
	\end{bmatrix} \, .
}
is parametrized by means of the three Euler angles $ a,\,b,\,c $. While the first orientation is defined by $\{a,b,c\}=\{0,0,0\}$ (i.e., $ \F\pl_\mathrm{init} =\mat{1}$), the second one corresponds to $ \{a,b,c\}=\{\pi/6,\pi/4,0\}$. 

\subsubsection{Comparison of the different algorithms}

\begin{figure}[ht]
  \centering
  \begin{minipage}[t]{0.02\textwidth}
    \vspace{45pt}
    \centering
    \rotatebox{90}{$\tau_{12}$ [GPa]}
  \end{minipage}
  \begin{minipage}[t]{0.45\textwidth}
    \vspace{0pt}
    \centering
    \begin{tikzpicture}
      \begin{axis}[
        width = 1.\linewidth, 
        height = 0.8\linewidth,
        xmin =  0.0, 
        xmax =  4.0, 
        ymin =  0.0, 
        ymax =  0.15, 
        xtick = {0.0, 1.0, 2.0, 3.0, 4.0},
        ytick = {0.0, 0.05, 0.1, 0.15},
        y tick label style={/pgf/number format/fixed},
        x tick label style={/pgf/number format/fixed},
        scaled ticks=false, 
        xlabel = {$F_{12}$},
        legend cell align=left,
        legend pos=south east
        ]
        \addplot[color = blue, 
                 mark = \empty,
                 line width = 1.2pt]
                 table {./Timeconv_coarse_shear_RM_200_000000.txt};
        \addplot[color = green,
                 mark = \empty,
                 dashed,
                 line width = 1.2pt]
                 table {./Timeconv_coarse_shear_RM_2000_000000.txt};
        \addplot[color = red,
                 mark = \empty,
                 dash dot,
                 line width = 1.2pt]
                 table {./Timeconv_coarse_shear_RM_20000_000000.txt};
        \legend{$ \Delta F_{12}= \num{2e-2} $, $ \Delta F_{12}= \num{2e-3} $, $ \Delta F_{12}= \num{2e-4} $}
      \end{axis}
    \end{tikzpicture}
    \subcaption{$ \{a,b,c\}=\{0,0,0\}$; Backward-Euler.}
  \end{minipage}
  \begin{minipage}[t]{0.45\textwidth}
    \vspace{0pt}
    \centering
    \begin{tikzpicture}
    \begin{axis}[
    width = 1.\linewidth, 
    height = 0.8\linewidth,
    xmin =  0.0, 
    xmax =  4.0, 
    ymin =  0.0, 
    ymax =  0.15, 
    xtick = {0.0, 1.0, 2.0, 3.0, 4.0},
    ytick = {0.0, 0.05, 0.1, 0.15},
    y tick label style={/pgf/number format/fixed},
    x tick label style={/pgf/number format/fixed},
    scaled ticks=false, 
    xlabel = {$F_{12}$},
    legend cell align=left,
    legend pos=south east
    ]
    \addplot[color = blue, 
    mark = \empty,
    line width = 1.2pt]
    table {./Timeconv_coarse_shear_RMExp_200_000000.txt};
    \addplot[color = green,
    mark = \empty,
    dashed,
    line width = 1.2pt]
    table {./Timeconv_coarse_shear_RMExp_2000_000000.txt};
    \addplot[color = red,
    mark = \empty,
    dash dot,
    line width = 1.2pt]
    table {./Timeconv_coarse_shear_RMExp_20000_000000.txt};
    \legend{$ \Delta F_{12}= \num{2e-2} $, $ \Delta F_{12}= \num{2e-3} $, $ \Delta F_{12}= \num{2e-4} $}
    \end{axis}
    \end{tikzpicture}
    \subcaption{$ \{a,b,c\}=\{0,0,0\}$; Exponential Map.}
  \end{minipage}%
\\
  \begin{minipage}[t]{0.02\textwidth}
    \vspace{45pt}
    \centering
    \rotatebox{90}{$\tau_{12}$ [GPa]}
  \end{minipage}
  \begin{minipage}[t]{0.45\textwidth}
    \vspace{0pt}
    \centering
    \begin{tikzpicture}
    \begin{axis}[
    width = \linewidth, 
    height = 0.8\linewidth,
    xmin =  0.0, 
    xmax =  4.0, 
    ymin =  0.0, 
    ymax =  0.12,
    xtick = {0.0, 1.0, 2.0, 3.0, 4.0},
    ytick = {0.0, 0.05, 0.1, 0.12},
    x tick label style={/pgf/number format/fixed},
    y tick label style={/pgf/number format/fixed},
    scaled ticks=true,
    xlabel = {$F_{12}$},
    legend cell align=left,
    legend pos=south east
    ]
    \addplot[color = blue, 
    mark = \empty,
    line width = 1.2pt]
    table {./Timeconv_coarse_shear_RMBE_200_004530.txt};
    \addplot[color = green,
    mark = \empty,
    dashed,
    line width = 1.2pt]
    table {./Timeconv_coarse_shear_RMBE_2000_004530.txt};
    \addplot[color = red,
    mark = \empty,
    dash dot,
    line width = 1.2pt]
    table {./Timeconv_coarse_shear_RMBE_20000_004530.txt};
    \legend{$ \Delta F_{12}= \num{2e-2} $, $ \Delta F_{12}= \num{2e-3} $, $ \Delta F_{12}= \num{2e-4} $}
    \end{axis}
    \end{tikzpicture}
    \subcaption{$ \{a,b,c\}=\{\pi/6,\pi/4,0\}$; Backward-Euler.}
  \end{minipage}%
  \begin{minipage}[t]{0.45\textwidth}
    \vspace{0pt}
    \centering
    \begin{tikzpicture}
    \begin{axis}[
    width = \linewidth, 
    height = .8\linewidth,
    xmin =  0.0, 
    xmax =  4.0, 
    ymin =  0.0, 
    ymax =  0.12, 
    xtick = {0.0, 1.0, 2.0, 3.0, 4.0},
    ytick = {0.0, 0.05, 0.1, 0.12},
    x tick label style={/pgf/number format/fixed},
    y tick label style={/pgf/number format/fixed},
    scaled ticks=true,
    xlabel = {$F_{12}$},
    legend cell align=left,
    legend pos=south east
    ]
    \addplot[color = blue, 
    mark = \empty,
    line width = 1.2pt]
    table {./Timeconv_coarse_shear_RMExp_200_004530.txt};
    \addplot[color = green,
    mark = \empty,
    dashed,
    line width = 1.2pt]
    table {./Timeconv_coarse_shear_RMExp_2000_004530.txt};
    \addplot[color = red,
    mark = \empty,
    dash dot,
    line width = 1.2pt]
    table {./Timeconv_coarse_shear_RMExp_20000_004530.txt};
    \legend{$ \Delta F_{12}= \num{2e-2} $, $ \Delta F_{12}= \num{2e-3} $, $ \Delta F_{12}= \num{2e-4} $}
    \end{axis}
    \end{tikzpicture}%
    \subcaption{$ \{a,b,c\}=\{\pi/6,\pi/4,0\}$; Exponential Map.}
  \end{minipage}%
  \caption{FCC crystal subjected to simple shear. Computations based on the return-mapping scheme combined with the Fischer-Burmeister NCP function, cf. \cite{FohrmeisterDiazMosler_2018a}. Computed stresses (coordinate $\tau_{12} $ of the Kirchhoff stress tensor) vs. shear strain $F_{12}$. Upper row: $\{0,0,0\}$ orientation of the FCC crystal. Lower row: $ \{\pi/6,\pi/4,0\}$ orientation of the crystal. Left column: algorithm based on the backward-Euler integration. Right column: algorithm based on an exponential time integration.}
  \label{fig: shear_RM_FB}
\end{figure}
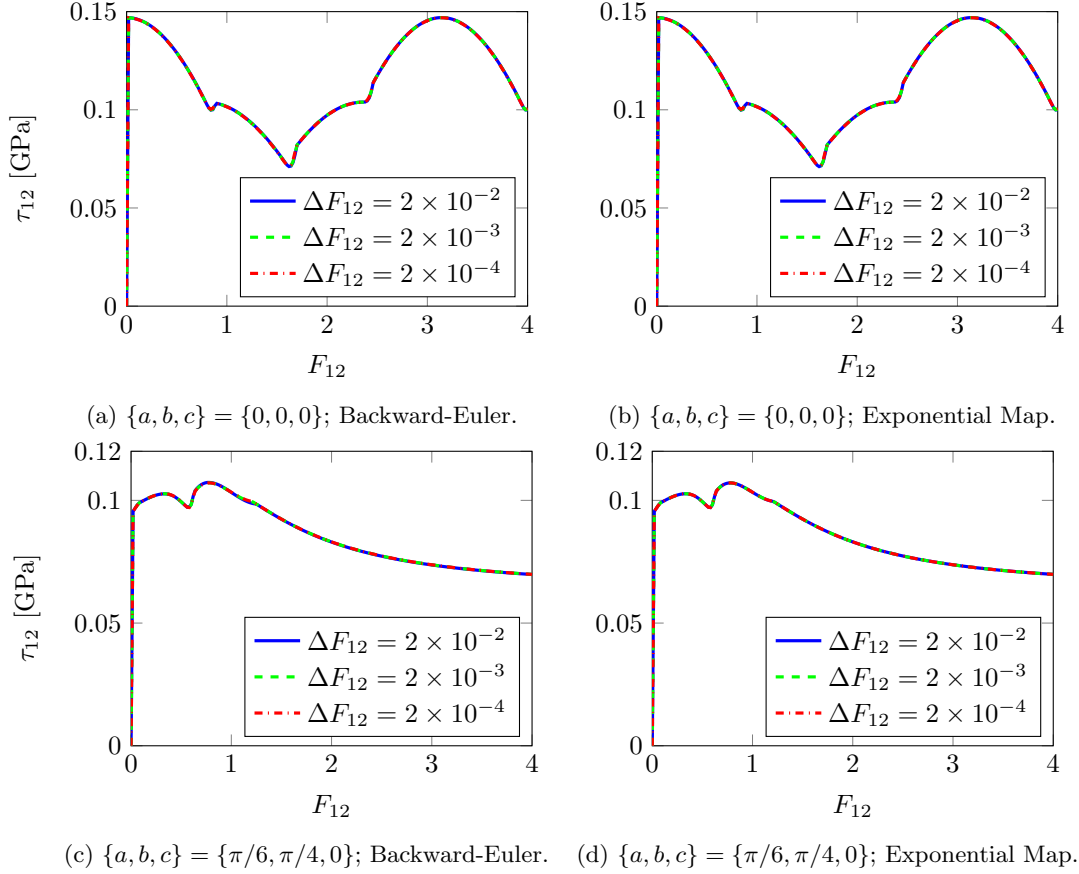

\begin{figure}[ht]
  \centering
  \begin{minipage}[t]{0.02\textwidth}
    \vspace{45pt}
    \centering
    \rotatebox{90}{$\tau_{12}$ [GPa]}
  \end{minipage}
  \begin{minipage}[t]{0.45\textwidth}
    \vspace{0pt}
    \centering
    \begin{tikzpicture}
    \begin{axis}[
        width = 1.\linewidth, 
        height = 0.8\linewidth,
        xmin =  0.0, 
        xmax =  4.0, 
        ymin =  0.0, 
        ymax =  0.15, 
        xtick = {0.0, 1.0, 2.0, 3.0, 4.0},
        ytick = {0.0, 0.05, 0.1, 0.15},
        y tick label style={/pgf/number format/fixed},
        x tick label style={/pgf/number format/fixed},
        scaled ticks=false, 
        xlabel = {$F_{12}$},
        legend cell align=left,
        legend pos=south east
        ]
        \addplot[color = blue, 
        mark = \empty,
        line width = 1.2pt]
        table {./Timeconv_coarse_shear_Var_200_000000.txt};
        \addplot[color = green,
        mark = \empty,
        dashed,
        line width = 1.2pt]
        table {./Timeconv_coarse_shear_Var_2000_000000.txt};
        \addplot[color = red,
        mark = \empty,
        dash dot,
        line width = 1.2pt]
        table {./Timeconv_coarse_shear_Var_20000_000000.txt};
        \legend{$ \Delta F_{12}= \num{2e-2} $, $ \Delta F_{12}= \num{2e-3} $, $ \Delta F_{12}= \num{2e-4} $}
        \end{axis}
    \end{tikzpicture}
    \subcaption{$ \{a,b,c\}=\{0,0,0\}$; Backward-Euler.}
  \end{minipage}
  \begin{minipage}[t]{0.45\textwidth}
    \vspace{0pt}
    \centering
    \begin{tikzpicture}
        \begin{axis}[
        width = 1.\linewidth, 
        height = 0.8\linewidth,
        xmin =  0.0, 
        xmax =  4.0, 
        ymin =  0.0, 
        ymax =  0.15, 
        xtick = {0.0, 1.0, 2.0, 3.0, 4.0},
        ytick = {0.0, 0.05, 0.1, 0.15},
        y tick label style={/pgf/number format/fixed},
        x tick label style={/pgf/number format/fixed},
        scaled ticks=false, 
        xlabel = {$F_{12}$},
        legend cell align=left,
        legend pos=south east
        ]
        \addplot[color = blue, 
        mark = \empty,
        line width = 1.2pt]
        table {./Timeconv_coarse_shear_VarExp_200_000000.txt};
        \addplot[color = green,
        mark = \empty,
        dashed,
        line width = 1.2pt]
        table {./Timeconv_coarse_shear_VarExp_2000_000000.txt};
        \addplot[color = red,
        mark = \empty,
        dash dot,
        line width = 1.2pt]
        table {./Timeconv_coarse_shear_VarExp_20000_000000.txt};
        \legend{$ \Delta F_{12}= \num{2e-2} $, $ \Delta F_{12}= \num{2e-3} $, $ \Delta F_{12}= \num{2e-4} $}
        \end{axis}
    \end{tikzpicture}
    \subcaption{$ \{a,b,c\}=\{0,0,0\}$; Exponential Map.}
  \end{minipage}%
\\
  \begin{minipage}[t]{0.02\textwidth}
    \vspace{45pt}
    \centering
    \rotatebox{90}{$\tau_{12}$ [GPa]}
  \end{minipage}
  \begin{minipage}[t]{0.45\textwidth}
    \vspace{0pt}
    \centering
    \begin{tikzpicture}
    \begin{axis}[
    width = \linewidth, 
    height = 0.8\linewidth,
    xmin =  0.0, 
    xmax =  4.0, 
    ymin =  0.0, 
    ymax =  0.12, 
    xtick = {0.0, 1.0, 2.0, 3.0, 4.0},
    ytick = {0.0, 0.05, 0.1, 0.12},
    x tick label style={/pgf/number format/fixed},
    y tick label style={/pgf/number format/fixed},
    scaled ticks=false, 
    xlabel = {$F_{12}$},
    legend cell align=left,
    legend pos=south east
    ]
    \addplot[color = blue, 
    mark = \empty,
    line width = 1.2pt]
    table {./Timeconv_coarse_shear_VarBE_200_004530.txt};
    \addplot[color = green,
    mark = \empty,
    dashed,
    line width = 1.2pt]
    table {./Timeconv_coarse_shear_VarBE_2000_004530.txt};
    \addplot[color = red,
    mark = \empty,
    dash dot,
    line width = 1.2pt]
    table {./Timeconv_coarse_shear_VarBE_20000_004530.txt};
    \legend{$ \Delta F_{12}= \num{2e-2} $, $ \Delta F_{12}= \num{2e-3} $, $ \Delta F_{12}= \num{2e-4} $}
    \end{axis}
    \end{tikzpicture}
    \subcaption{$ \{a,b,c\}=\{\pi/6,\pi/4,0\}$; Backward-Euler.}
  \end{minipage}%
  \begin{minipage}[t]{0.45\textwidth}
    \vspace{0pt}
    \centering
    \begin{tikzpicture}
    \begin{axis}[
    width = \linewidth, 
    height = .8\linewidth,
    xmin =  0.0, 
    xmax =  4.0, 
    ymin =  0.0, 
    ymax =  0.12, 
    xtick = {0.0, 1.0, 2.0, 3.0, 4.0},
    ytick = {0.0, 0.05, 0.1, 0.12},
    x tick label style={/pgf/number format/fixed},
    y tick label style={/pgf/number format/fixed},
    scaled ticks=false, 
    xlabel = {$F_{12}$},
    legend cell align=left,
    legend pos=south east
    ]
    \addplot[color = blue, 
    mark = \empty,
    line width = 1.2pt]
    table {./Timeconv_coarse_shear_VarExp_200_004530.txt};
    \addplot[color = green,
    mark = \empty,
    dashed,
    line width = 1.2pt]
    table {./Timeconv_coarse_shear_VarExp_2000_004530.txt};
    \addplot[color = red,
    mark = \empty,
    dash dot,
    line width = 1.2pt]
    table {./Timeconv_coarse_shear_VarExp_20000_004530.txt};
    \legend{$ \Delta F_{12}= \num{2e-2} $, $ \Delta F_{12}= \num{2e-3} $, $ \Delta F_{12}= \num{2e-4} $}
    \end{axis}
    \end{tikzpicture}
    \subcaption{$ \{a,b,c\}=\{\pi/6,\pi/4,0\}$; Exponential Map.}
  \end{minipage}%
  \caption{FCC crystal subjected to simple shear. Computations utilizing the novel variational algorithm based on the nonlinear complementary problem, see Subsection~\ref{subsec:Methods-FischerBurmeister}. Computed stresses (coordinate $\tau_{12} $ of the Kirchhoff stress tensor) vs. shear strain $F_{12}$. Upper row: $\{0,0,0\}$ orientation of the FCC crystal. Lower row: $ \{\pi/6,\pi/4,0\}$ orientation of the crystal. Left column: algorithm based on the backward-Euler integration. Right column: algorithm based on an exponential time integration.}
  \label{fig: shear_var_FB}
\end{figure}
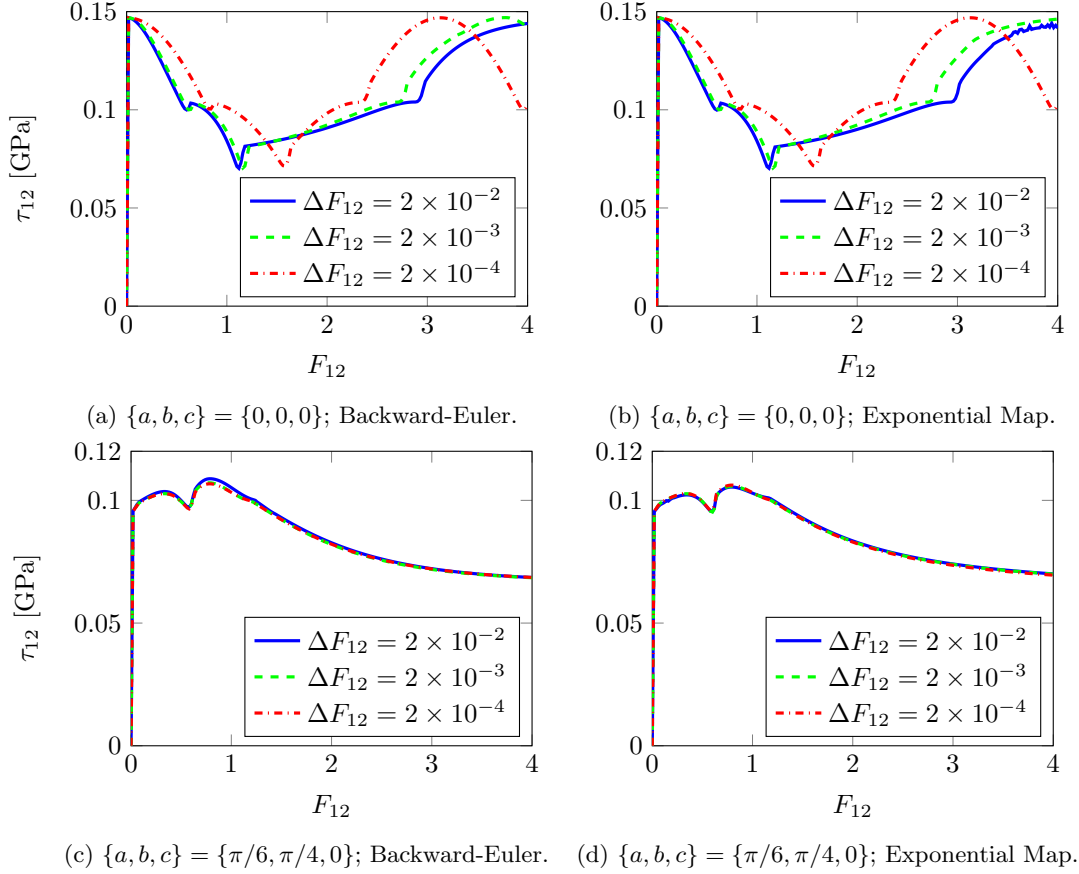

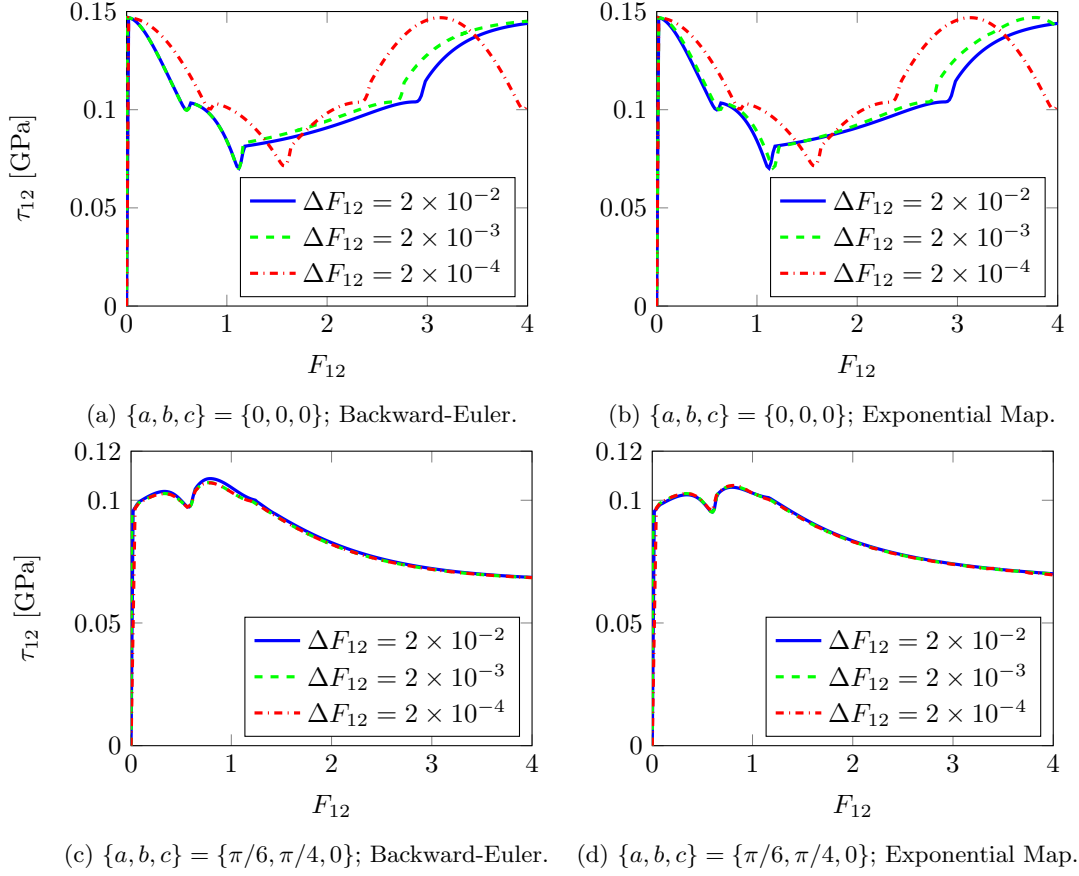
\begin{figure}[ht]
  \centering
  \begin{minipage}[t]{0.02\textwidth}
    \vspace{45pt}
    \centering
    \rotatebox{90}{$\tau_{12}$ [GPa]}
  \end{minipage}
  \begin{minipage}[t]{0.45\textwidth}
    \vspace{0pt}
    \centering
    \begin{tikzpicture}
    \begin{axis}[
        width = 1.\linewidth, 
        height = 0.8\linewidth,
        xmin =  0.0, 
        xmax =  4.0, 
        ymin =  0.0, 
        ymax =  0.15, 
        xtick = {0.0, 1.0, 2.0, 3.0, 4.0},
        ytick = {0.0, 0.05, 0.1, 0.15},
        y tick label style={/pgf/number format/fixed},
        x tick label style={/pgf/number format/fixed},
        scaled ticks=false, 
        xlabel = {$F_{12}$},
        legend cell align=left,
        legend pos=south east
        ]
        \addplot[color = blue, 
        mark = \empty,
        line width = 1.2pt]
        table {./Timeconv_coarse_shear_VarBE_AugL_200_000000.txt};
        \addplot[color = green,
        mark = \empty,
        dashed,
        line width = 1.2pt]
        table {./Timeconv_coarse_shear_VarBE_AugL_2000b_000000.txt};
        \addplot[color = red,
        mark = \empty,
        dash dot,
        line width = 1.2pt]
        table {./Timeconv_coarse_shear_VarBE_AugL_20000_000000.txt};
        \legend{$ \Delta F_{12}= \num{2e-2} $, $ \Delta F_{12}= \num{2e-3} $, $ \Delta F_{12}= \num{2e-4} $}
        \end{axis}
    \end{tikzpicture}
    \subcaption{$ \{a,b,c\}=\{0,0,0\}$; Backward-Euler.}
  \end{minipage}
  \begin{minipage}[t]{0.45\textwidth}
    \vspace{0pt}
    \centering
    \begin{tikzpicture}
        \begin{axis}[
        width = 1.\linewidth, 
        height = 0.8\linewidth,
        xmin =  0.0, 
        xmax =  4.0, 
        ymin =  0.0, 
        ymax =  0.15, 
        xtick = {0.0, 1.0, 2.0, 3.0, 4.0},
        ytick = {0.0, 0.05, 0.1, 0.15},
        y tick label style={/pgf/number format/fixed},
        x tick label style={/pgf/number format/fixed},
        scaled ticks=false, 
        xlabel = {$F_{12}$},
        legend cell align=left,
        legend pos=south east
        ]
        \addplot[color = blue, 
        mark = \empty,
        line width = 1.2pt]
        table {./Timeconv_coarse_shear_VarExp_AugL_200_000000.txt};
        \addplot[color = green,
        mark = \empty,
        dashed,
        line width = 1.2pt]
        table {./Timeconv_coarse_shear_VarExp_AugL_2000_000000.txt};
        \addplot[color = red,
        mark = \empty,
        dash dot,
        line width = 1.2pt]
        table {./Timeconv_coarse_shear_VarExp_AugL_20000_000000.txt};
        \legend{$ \Delta F_{12}= \num{2e-2} $, $ \Delta F_{12}= \num{2e-3} $, $ \Delta F_{12}= \num{2e-4} $}
        \end{axis}
    \end{tikzpicture}
    \subcaption{$ \{a,b,c\}=\{0,0,0\}$; Exponential Map.}
  \end{minipage}%
\\
  \begin{minipage}[t]{0.02\textwidth}
    \vspace{45pt}
    \centering
    \rotatebox{90}{$\tau_{12}$ [GPa]}
  \end{minipage}
  \begin{minipage}[t]{0.45\textwidth}
    \vspace{0pt}
    \centering
    \begin{tikzpicture}
    \begin{axis}[
    width = \linewidth, 
    height = 0.8\linewidth,
    xmin =  0.0, 
    xmax =  4.0, 
    ymin =  0.0, 
    ymax =  0.12, 
    xtick = {0.0, 1.0, 2.0, 3.0, 4.0},
    ytick = {0.0, 0.05, 0.1, 0.12},
    x tick label style={/pgf/number format/fixed},
    y tick label style={/pgf/number format/fixed},
    scaled ticks=false, 
    xlabel = {$F_{12}$},
    legend cell align=left,
    legend pos=south east
    ]
    \addplot[color = blue, 
    mark = \empty,
    line width = 1.2pt]
    table {./Timeconv_coarse_shear_VarBE_AugL_200_004530.txt};
    \addplot[color = green,
    mark = \empty,
    dashed,
    line width = 1.2pt]
    table {./Timeconv_coarse_shear_VarBE_AugL_2000_004530.txt};
    \addplot[color = red,
    mark = \empty,
    dash dot,
    line width = 1.2pt]
    table {./Timeconv_coarse_shear_VarBE_AugL_20000_004530.txt};
    \legend{$ \Delta F_{12}= \num{2e-2} $, $ \Delta F_{12}= \num{2e-3} $, $ \Delta F_{12}= \num{2e-4} $}
    \end{axis}
    \end{tikzpicture}
    \subcaption{$ \{a,b,c\}=\{\pi/6,\pi/4,0\}$; Backward-Euler.}
  \end{minipage}%
  \begin{minipage}[t]{0.45\textwidth}
    \vspace{0pt}
    \centering
    \begin{tikzpicture}
    \begin{axis}[
    width = \linewidth, 
    height = .8\linewidth,
    xmin =  0.0, 
    xmax =  4.0, 
    ymin =  0.0, 
    ymax =  0.12, 
    xtick = {0.0, 1.0, 2.0, 3.0, 4.0},
    ytick = {0.0, 0.05, 0.1, 0.12},
    x tick label style={/pgf/number format/fixed},
    y tick label style={/pgf/number format/fixed},
    scaled ticks=false, 
    xlabel = {$F_{12}$},
    legend cell align=left,
    legend pos=south east
    ]
    \addplot[color = blue, 
    mark = \empty,
    line width = 1.2pt]
    table {./Timeconv_coarse_shear_VarEXP_AugL_200_004530.txt};
    \addplot[color = green,
    mark = \empty,
    dashed,
    line width = 1.2pt]
    table {./Timeconv_coarse_shear_VarEXP_AugL_2000_004530.txt};
    \addplot[color = red,
    mark = \empty,
    dash dot,
    line width = 1.2pt]
    table {./Timeconv_coarse_shear_VarEXP_AugL_20000_004530.txt};
    \legend{$ \Delta F_{12}= \num{2e-2} $, $ \Delta F_{12}= \num{2e-3} $, $ \Delta F_{12}= \num{2e-4} $}
    \end{axis}
    \end{tikzpicture}
    \subcaption{$ \{a,b,c\}=\{\pi/6,\pi/4,0\}$; Exponential Map.}
  \end{minipage}%
  \caption{FCC crystal subjected to simple shear. Computations based on the variational augmented Lagrangian formulation, see Subsection~ \ref{subsec:Methods-AugLagrange}. Computed stresses (coordinate $\tau_{12} $ of the Kirchhoff stress tensor) vs. shear strain $F_{12}$. Upper row: $\{0,0,0\}$ orientation of the FCC crystal. Lower row: $ \{\pi/6,\pi/4,0\}$ orientation of the crystal. Left column: algorithm based on the backward-Euler integration. Right column: algorithm based on an exponential time integration.}
  \label{fig: shear_var_AugL}
\end{figure}

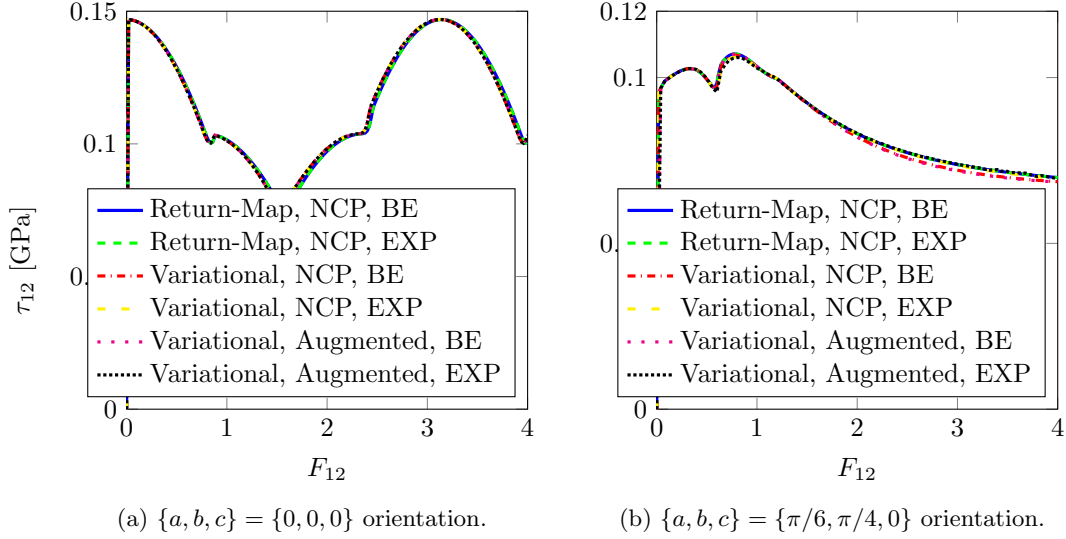
\begin{figure}[ht]
  \centering
  \begin{minipage}[t]{0.02\textwidth}
    \vspace{80pt}
    \centering
    \rotatebox{90}{$\tau_{12}$ [GPa]}
  \end{minipage}
  \begin{minipage}[t]{0.45\textwidth}
    \vspace{0pt}
    \centering
    \begin{tikzpicture}
      \begin{axis}[
        width = 1.\linewidth, 
        height = 1.\linewidth,
        xmin =  0.0, 
        xmax =  4.0, 
        ymin =  0.0, 
        ymax =  0.15, 
        xtick = {0.0, 1.0, 2.0, 3.0, 4.0},
        ytick = {0.0, 0.05, 0.1, 0.15},
        y tick label style={/pgf/number format/fixed},
        x tick label style={/pgf/number format/fixed},
        scaled ticks=false, 
        xlabel = {$F_{12}$},
        legend cell align=left,
        legend pos=south east
        ]
        \addplot[color = blue, 
                 mark = \empty,
                 line width = 1.2pt]
                 table {./Timeconv_coarse_shear_RM_20000_000000.txt};
        \addplot[color = green,
                 mark = \empty,
                 dashed,
                 line width = 1.2pt]
                 table {./Timeconv_coarse_shear_RMExp_20000_000000.txt};
        \addplot[color = red,
                 mark = \empty,
                 dash dot,
                 line width = 1.2pt]
                 table {./Timeconv_coarse_shear_Var_20000_000000.txt};
        \addplot[color = yellow,
                 mark = \empty,
                 loosely dashed,
                 line width = 1.2pt]
                 table {./Timeconv_coarse_shear_VarExp_20000_000000.txt};
        \addplot[color = magenta,
                mark = \empty,
                loosely dotted,
                line width = 1.2pt]
                table {./Timeconv_coarse_shear_VarBE_AugL_20000_000000.txt};
        \addplot[color = black,
                mark = \empty,
                densely dotted,
                line width = 1.2pt]
                table {./Timeconv_coarse_shear_VarExp_AugL_20000_000000.txt};
        \legend{{Return-Map, NCP, BE}, 
            {Return-Map, NCP, EXP},
            {Variational, NCP, BE},
            {Variational, NCP, EXP},
            {Variational, Augmented, BE},
            {Variational, Augmented, EXP}}
      \end{axis}
    \end{tikzpicture}
    \subcaption{$ \{a,b,c\}=\{0,0,0\}$ orientation.}
  \end{minipage}
  \begin{minipage}[t]{0.45\textwidth}
        \vspace{0pt}
        \centering
        \begin{tikzpicture}
        \begin{axis}[
        width = 1.\linewidth, 
        height =1.\linewidth,
        xmin =  0.0, 
        xmax =  4.0, 
        ymin =  0.0, 
        ymax =  0.12, 
        xtick = {0.0, 1.0, 2.0, 3.0, 4.0},
        ytick = {0.0, 0.05, 0.1, 0.12},
        y tick label style={/pgf/number format/fixed},
        x tick label style={/pgf/number format/fixed},
        scaled ticks=false, 
        xlabel = {$F_{12}$},
        legend cell align=left,
        legend pos=south east
        ]
        \addplot[color = blue, 
        mark = \empty,
        line width = 1.2pt]
        table {./Timeconv_coarse_shear_RMBE_20000_004530.txt};
        \addplot[color = green,
        mark = \empty,
        dashed,
        line width = 1.2pt]
        table {./Timeconv_coarse_shear_RMExp_20000_004530.txt};
        \addplot[color = red,
        mark = \empty,
        dash dot,
        line width = 1.2pt]
        table {./Timeconv_coarse_shear_VarBE_20000_004530.txt};
        \addplot[color = yellow,
        mark = \empty,
        loosely dashed,
        line width = 1.2pt]
        table {./Timeconv_coarse_shear_VarExp_20000_004530.txt};
        \addplot[color = magenta,
        mark = \empty,
        loosely dotted,
        line width = 1.2pt]
        table {./Timeconv_coarse_shear_VarBE_AugL_20000_004530.txt};
        \addplot[color = black,
        mark = \empty,
        densely dotted,
        line width = 1.2pt]
        table {./Timeconv_coarse_shear_VarExp_AugL_20000_004530.txt};
        \legend{{Return-Map, NCP, BE}, 
        	{Return-Map, NCP, EXP},
        	{Variational, NCP, BE},
        	{Variational, NCP, EXP},
        	{Variational, Augmented, BE},
        	{Variational, Augmented, EXP}}
        \end{axis}
        \end{tikzpicture}
        \subcaption{$ \{a,b,c\}=\{\pi/6,\pi/4,0\}$ orientation.}
    \end{minipage}
  \caption{FCC crystal subjected to simple shear. Comparison of the three different algorithms: return-mapping scheme combined with the Fischer-Burmeister NCP function (Return-Map, NCP), variational algorithm based on the Fischer-Burmeister NCP function (Variational, NCP) and variational algorithm based on the augmented Lagrangian formulation. Computed stresses (coordinate $\tau_{12} $ of the Kirchhoff stress tensor) vs. shear strain $F_{12}$. Left row: $\{0,0,0\}$ orientation of the FCC crystal. Right row: $\{\pi/6,\pi/4,0\}$ orientation of the FCC crystal. Within all computations, the finest time discretization with $\Delta F_{12}= \num{2e-4}$ was chosen.}
  \label{fig: overview_20000}
\end{figure}

The numerically predicted stress-strain response is summarized in Fig.~\ref{fig: shear_RM_FB} --- Fig.~\ref{fig: overview_20000}. Within all diagrams, the $\tau_{12}$ coordinate of the Kirchhoff stress tensor is shown as a function in terms of the prescribed shear strain $F_{12}$. Fig.~\ref{fig: shear_RM_FB} corresponds to the return-mapping scheme combined with the Fischer-Burmeister NCP function -- including the novel rescaling technique.
%
According to Fig.~\ref{fig: shear_RM_FB} the predicted mechanical response is almost independent with respect to the employed time discretization. To be more precise, the results neither depend on the time integration scheme (backward-Euler integration vs. exponential mapping), nor on the size of the loading increment ($\Delta F_{12}$). Even the coarsest time discretization ($\Delta F_{12}=2\times 10^{-2}$) predicts the same response as the finest time discretization ($\Delta F_{12}=2\times 10^{-4}$). Finally, it is noted that the results associated with the different crystal orientations agree with those reported in \cite{Schmidt-Baldassari:03,Prger2020ACS}.

The results obtained from the novel variational algorithm based on the Fischer-Burmeister NCP function (see Subsection~\ref{subsec:Methods-FischerBurmeister}) are given in Fig.~\ref{fig: shear_var_FB}.
%
It can be seen that the algorithm performs similar to the return-mapping scheme combined with the NCP function for orientation $ \{a,b,c\}=\{\pi/6,\pi/4,0\}$. By way of contrast and independent of the employed time integration scheme, the predicted stress-strain response depends on the size of the loading amplitude $\Delta F_{12}$ for orientation $\{a,b,c\}=\{0,0,0\}$. Only if the time discretization is sufficiently fine ($\Delta F_{12}=2\times 10^{-4}$), all algorithms lead to an identical mechanical response, cf. Fig.~\ref{fig: overview_20000}. It bears emphasis though that this observation cannot be explained only by means of a classic convergence behavior. To be more precise and as already mentioned before, rate-independent crystal plasticity theory often allows for several admissible solutions and up to know it is an open question which of these admissible solutions is chosen by a certain algorithm, cf. \cite{Niehueser_Mosler:23}. This question shall be answered in a separate paper. Furthermore, it also should be recalled that variational updates only approximate the yield function, cf. Eq.~\eqref{eq-ex-dIdlamda}. Only in the limiting case $\Delta F_{12}\to 0$, the respective error vanishes.

Fig.~\ref{fig: shear_var_AugL} shows the results as predicted by the second novel variational update, i.e., the algorithm based on the augmented Lagrangian formulation, cf. Subsection~\ref{subsec:Methods-AugLagrange}.
%
By comparing Fig.~\ref{fig: shear_var_AugL} to Fig.~\ref{fig: shear_var_FB} one can see that both novel variational lead to similar results. Particularly, the influence of the loading amplitude $F_{12}$ on the stress-strain diagram can again be observed for orientation $\{a,b,c\}=\{0,0,0\}$. Only in the limiting case $\Delta F_{12}\to 0$, all algorithms lead to an identical mechanical response, cf. Fig.~\ref{fig: overview_20000}.

For a better comparison of all three algorithms, i.e., (1) the return-mapping scheme combined with the Fischer-Burmeister NCP function, (2) the novel variational update based on the Fischer-Burmeister NCP function and (3) the novel variational update relying on the augmented Lagrangian formulation, all predicted stress-strain responses are summarized in Fig.~\ref{fig: overview_20000}. 
%
The results correspond to the finest temporal time discretization ($\Delta F_{12}=2\times 10^{-4}$). According to Fig.~\ref{fig: overview_20000}, all algorithms lead to the same mechanical response, if the time discretization is sufficiently small. It should be emphasized though that this finding cannot be universally valid due to the non-uniqueness of crystal plasticity theory. This point shall be analyzed in a forthcoming paper. Nevertheless, all algorithms indeed turned out to be numerically robust and efficient. Furthermore, due to the underlying variational structure, the novel updates show additional advantages such as a canonical fundament for error estimation and adaptivity. Finally it is noted that orientation $\{a,b,c\}=\{0,0,0\}$ represents an ideal benchmark for crystal plasticity theory, since the number of active slip systems is high and the resulting mechanical response is very complex and non-monotonic in time.

\subsubsection{Effect of the scaling parameter within the Fischer-Burmeister NCP approach}

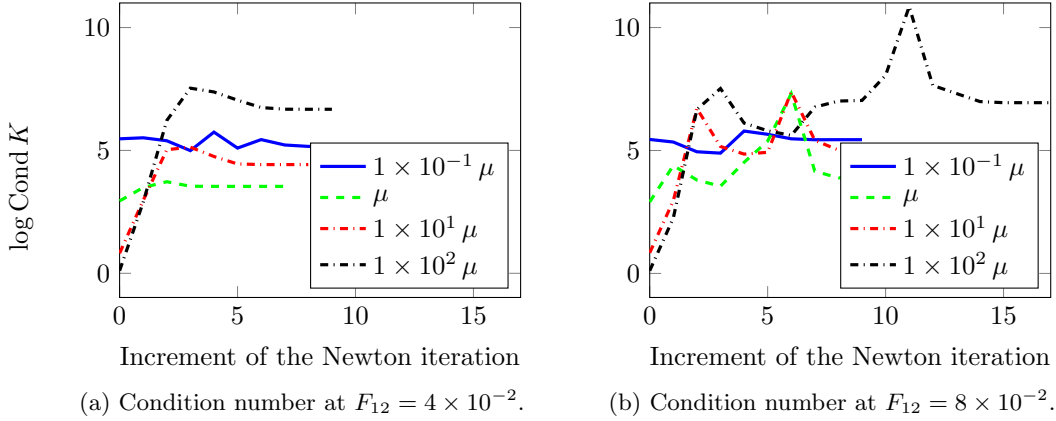
\begin{figure}[ht]
  \centering
  \begin{minipage}[t]{0.02\textwidth}
    \vspace{45pt}
    \centering
    \rotatebox{90}{$ \mathrm{log}\, \mathrm{Cond} \, K $}
  \end{minipage}
  \begin{minipage}[t]{0.45\textwidth}
    \vspace{0pt}
    \centering
    \begin{tikzpicture}
      \begin{axis}[
        width = 1.\linewidth, 
        height = 0.8\linewidth,
        xmin =  0, 
        xmax =  17, 
        ymax =  11, 
        x tick label style={/pgf/number format/fixed},
        y tick label style={/pgf/number format/fixed},
        scaled ticks=true,
        xlabel = {Increment of the Newton iteration},
        legend cell align=left,
        legend pos=south east
        ]
        \addplot[color = blue, 
             mark = \empty,
             line width = 1.2pt]
             table {./Condition_001_m1.txt};
        \addplot[color = green,
             mark = \empty,
             dashed,
             line width = 1.2pt]
             table {./Condition_001_0.txt};
        \addplot[color = red,
             mark = \empty,
             dash dot,
             line width = 1.2pt]
             table {./Condition_001_1.txt};
        \addplot[color = black,
             mark = \empty,
             dash dot,
             line width = 1.2pt]
             table {./Condition_001_2.txt};
        \legend{$ \num{1e-1}\,\mu $, $ \mu  $, $  \num{1e1}\,\mu $, $ \num{1e2}\,\mu  $}
      \end{axis}
    \end{tikzpicture}%
    \subcaption{Condition number at $ F_{12} = \num{4e-2} $.}
  \end{minipage}
  \begin{minipage}[t]{0.45\textwidth}
    \vspace{0pt}
    \centering
      \begin{tikzpicture}
          \begin{axis}[
          width = 1.\linewidth, 
          height = 0.8\linewidth,
          xmin =  0, 
          xmax =  17, 
          ymax =  11, 
          x tick label style={/pgf/number format/fixed},
          y tick label style={/pgf/number format/fixed},
          scaled ticks=true,
          xlabel = {Increment of the Newton iteration},
          legend cell align=left,
          legend pos=south east
          ]
          \addplot[color = blue, 
          mark = \empty,
          line width = 1.2pt]
          table {./Condition_002_m1.txt};
          \addplot[color = green,
          mark = \empty,
          dashed,
          line width = 1.2pt]
          table {./Condition_002_0.txt};
          \addplot[color = red,
          mark = \empty,
          dash dot,
          line width = 1.2pt]
          table {./Condition_002_1.txt};
          \addplot[color = black,
          mark = \empty,
          dash dot,
          line width = 1.2pt]
          table {./Condition_002_2.txt};
          \legend{$ \num{1e-1}\,\mu $, $ \mu  $, $  \num{1e1}\,\mu $, $ \num{1e2}\,\mu  $}
          \end{axis}
      \end{tikzpicture}
      \subcaption{Condition number at $ F_{12} = \num{8e-2} $.}
  \end{minipage}
  \caption{FCC crystal subjected to simple shear. Effect of scaling parameter $w$ within the modified Fischer-Burmeister function on condition number $K$ of tangent $\partial G\ssi/\partial \dlambda^{(j)} $ during a Newton iteration (condition number is based on the spectral norm). The results correspond to two different time steps ($ F_{12} = \num{4e-2} $ and $ F_{12} = \num{8e-2} $).}
  \label{fig:condition_scaling}
\end{figure}

The influence of scaling parameter $w$ within the Fischer-Burmeister function~\eqref{eq-FischerBurmeister-smoothed-212} is shown in Fig.~\ref{fig:condition_scaling}.  The results correspond to simulations performed by means of the variational algorithm based on a Backward-Euler time integration. Orientation $ \{a,b,c\}=\{\pi/6,\pi/4,0\}$ was considered. While the left diagram is associated with time step $ F_{12} = \num{4e-2} $, the diagram on the right hand side is related to $ F_{12} = \num{8e-2} $. Both snapshots have been extracted from the same simulation where a temporal discretization with $\Delta F_{12} = \num{2e-2} $ was chosen. Fig.~\ref{fig:condition_scaling} displays the evolution of the condition number of matrix $\partial G\ssi/\partial \dlambda^{(j)} $ within the (undamped) Newton iteration.
%

It can be seen, that the condition number is small within the first Newton iterations, if scaling parameter $w$ is sufficiently large. Subsequently, the condition number increases during the Newton iteration. This behavior is in line with the mathematical analysis in Subsection~\ref{sec-variational-ncp-gradient}. Equally important, the condition number does not tend to infinity, if convergence is obtained (last iteration step within the Newton iteration). As explained in Subsection~\ref{sec-variational-ncp-gradient}, this positive feature is due to the regularization parameter which has been set to $\delta=1\times 10^{-10}$. Without this regularization ($\delta=0$), matrix $\partial G\ssi/\partial \dlambda^{(j)} $ would be singular and thus, the condition number would be infinite.

In addition to the aforementioned general trends, the dependence of the condition number on parameter $w$ is obvious. In this connection, $w$ has been varied within the range $[1\times 10^{-1}\,\mu,1\times 10^{2}\,\mu]$. For smaller values of $w$ --- again in line with the mathematical analysis in Subsection~\ref{sec-variational-ncp-gradient} --- no convergence was obtained. Analogously, too large values of $w$ also lead to numerical problems. Within the chosen range $w\in[1\times 10^{-1}\,\mu,1\times 10^{2}\,\mu]$, value $w=\mu$ is the best choice at load step $F_{12}=4\times 10^{-2}$. According to Fig.~\ref{fig:condition_scaling} (left), value $w=\mu$ leads to the smallest condition number (already after iteration step 2) and the smallest number of Newton iterations up to convergence. As a consequence, $w=\mu$ is numerically most robust (smallest condition number) and numerically most efficient (smallest number of Newton iterations). Interestingly, choice $w=\mu$ was already well motivated by analyzing the magnitudes of the terms defining the Fischer-Burmeister function, cf. page~\pageref{choice-of-w} in Subsection~\ref{choice-of-w}. Similar to load step $F_{12}=4\times 10^{-2}$, the performance of the algorithm is also very good at $F_{12}=8\times 10^{-2}$ for choice $w=\mu$ . To be more precise, $w=\mu$ again results in the smallest condition number within the vicinity of the converged state. Furthermore, $w=\mu$ is also characterized by a small number of Newton iterations up to convergence. However, choice $w=1\times 10^{-1}\,\mu$ requires two iterations less at $F_{12}=8\times 10^{-2}$.

In summary and in line with the mathematical analyses in Subsection~\ref{sec-variational-ncp-gradient}, the numerical experiments presented in this section support the recommendation of choice $w\approx \mu$. This choice is motivated by the resulting numerical robustness (smallest condition number) as well as by the numerical efficiency (smallest number of Newton iterations).

\subsection{Tensile test of a single crystal}\label{subsec:TensileTest}

\begin{figure}[ht]
    \begin{minipage}[t]{0.48\textwidth}
    	\centering
    	\includegraphics[width=\linewidth]{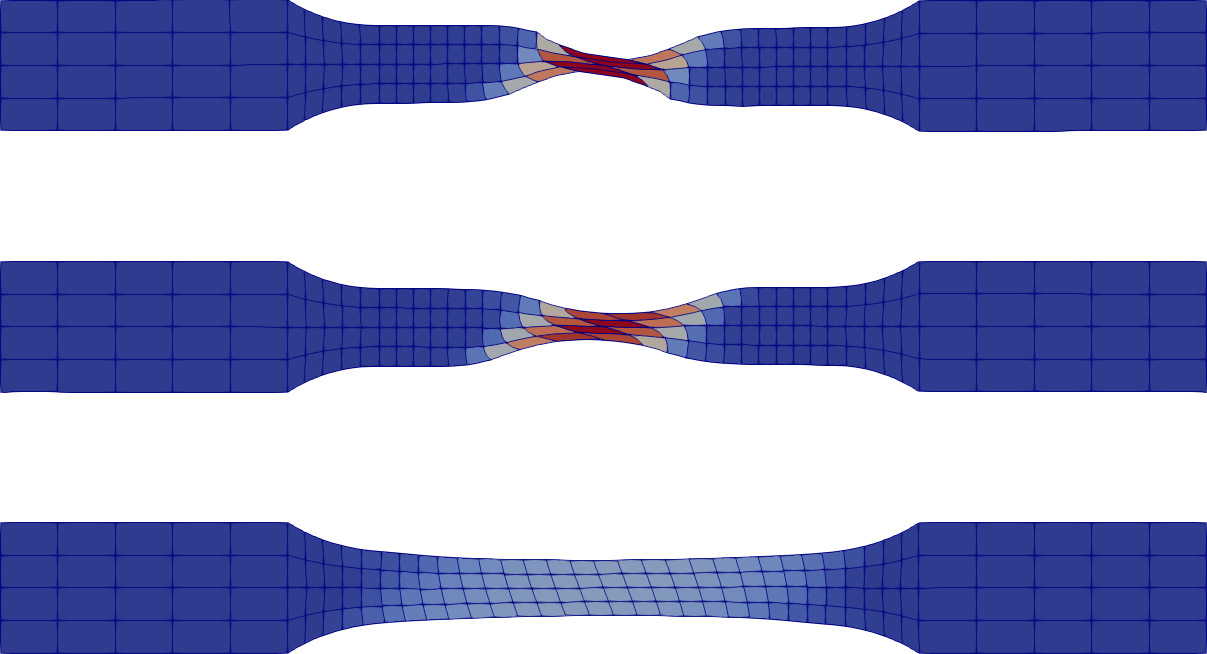}
        \\\vspace{2ex}
        \begin{tikzpicture}
        \node[anchor=south west,inner sep=0] (image) at (0,0) {\includegraphics[width=.5\linewidth]{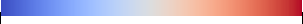}};
        \begin{scope}[x={(image.south east)},y={(image.north west)}]
            \node at (0,-.5) {$ 0 $};
            \node at (.5,-.5) {$ \| \vec{\alpha} \| $};
            \node at (1,-.5) {$ 2.3 $};
        \end{scope}
        \end{tikzpicture}
    	\subcaption{Distribution of accumulated plastic slip}
    \end{minipage}\hfill%
    \begin{minipage}[t]{0.48\textwidth}
        \centering
        \includegraphics[width=\linewidth]{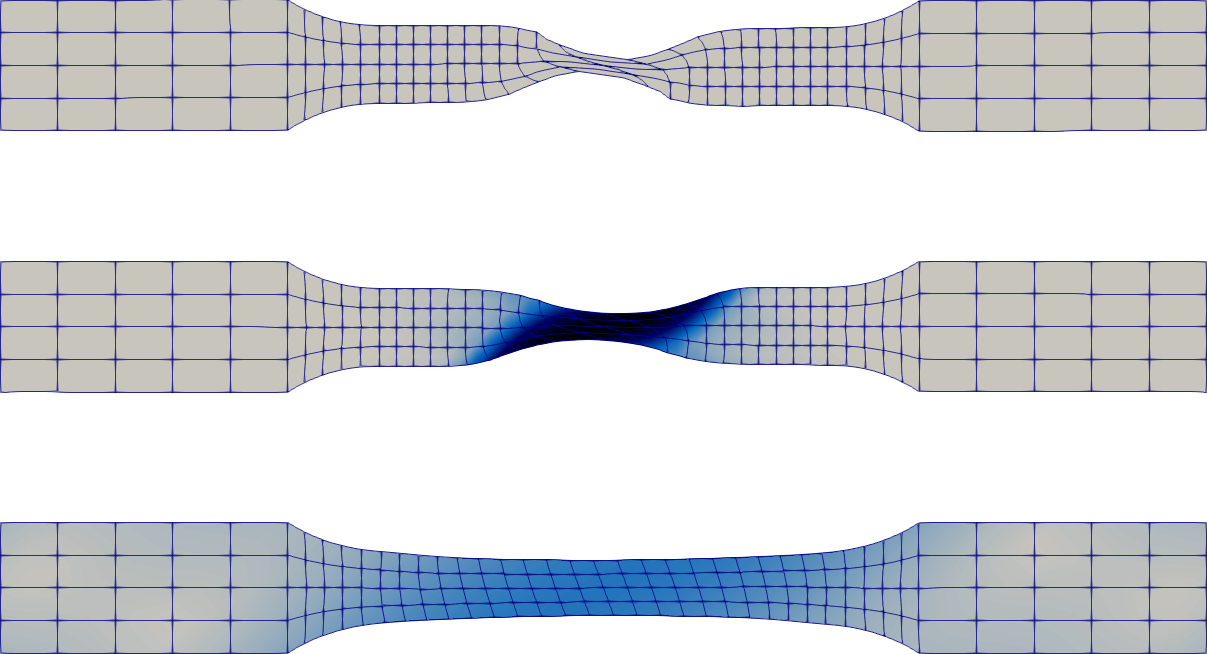}
        \\\vspace{2ex}
        \begin{tikzpicture}
        \node[anchor=south west,inner sep=0] (image) at (0,0) {\includegraphics[width=.5\linewidth]{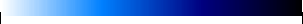}};
        \begin{scope}[x={(image.south east)},y={(image.north west)}]
            \node at (0,-.5) {$ 0 $};
            \node at (.5,-.5) {$ \| \vec{\varphi}^\mathrm{p} \| $};
            \node at (1,-.5) {$ 2 $};
        \end{scope}
        \end{tikzpicture}
        \subcaption{Distribution of norm of micromorphic field}
    \end{minipage}
    \caption{Tensile test of a single crystal. The computations are based on the novel variational update combined with the Fischer-Burmeister-function: (left) accumulated plastic slip ($ \alpha\ssi :=\sum_i\alpha^{(i)}$) and (right) norm of the micromorphic field ($ {\phip}\ssi :=\sum_i\varphi^{(i)}$) at the last time step at $ u = \SI{4.2}{\micro\metre} $ for different model parameters $c_2$. From top to bottom: $c_2=0$ (local model), $ c_2 = \SI{1e-6}{\mega\pascal\metre} $, $ c_2 = \SI{2}{\mega\pascal\metre} $.}
    \label{fig: Tensiletest_plot}
\end{figure}

Next, a tensile test of a single crystal is numerically analyzed. The geometry of the specimen represents a very downscaled version of a standard macroscopic tensile specimen with a total length of \SI{84}{\micro\metre} and a width of \SI{10}{\micro\metre}, see Fig.~\ref{fig: Tensiletest_plot}. Plane strain conditions are assumed. Plasticity is triggered by a geometric imperfection in the middle of the specimen: a tapered width to \SI{6}{\micro\metre}. Due to this imperfection and due to the shape of the specimen, the resulting stresses and strains are non-constant. For this reason, gradient hardening is active. A summary of the model parameters is given in Tab.~\ref{tab-matparam-tensile}
\begin{table}[ht!]
	\centering
	\begin{tabular}{|c|c|c|c|c|}
		\hline
		$\kappa$        &
		$\mu$           &
		$Q_0$         &
		$ c_1 $ &
		$ c_2 $ \\\hline\hline
\SI{49.98}{\giga\pascal} &
\SI{21.1}{\giga\pascal} &
\SI{6e-2}{\giga\pascal} &
\SI{0.1}{\giga\pascal} &
$0$ (local),\quad $\,\SI{1e-6}{\mega\pascal\metre}$,\quad$\SI{\,2}{\mega\pascal\metre}$   \\
		\hline
	\end{tabular}
	\caption{Tensile test of a single crystal: Model parameters. Perfect plasticity with gradient-hardening}
	\label{tab-matparam-tensile}
\end{table}
The crystal orientation is again chosen as $ \{a,b,c\}=\{\pi/6,\pi/4,0\}$ and thus, the crystal is not parallel to the axes of the specimen, such that geometrically asymmetric behavior of the boundary value problem is provoked.

The specimen has been analyzed numerically by means of the finite element method. For that purpose, a discretization with 160 8-noded serendipity quadrilateral elements is employed. In order to highlight the effect of gradient hardening, the simulations have been carried out by using different model parameters $ c_2 $. 

Results obtained from computations based on the novel variational update combined with the Fischer-Burmeister-function are shown in Fig.~\ref{fig: Tensiletest_plot}.
%
%

According to Fig.~\ref{fig: Tensiletest_plot}, the influence of gradient hardening parameter $c_2$ is obvious. While localized plastic deformations are predicted by the local model $c_2=0$, increasing $c_2$ leads to a smearing of the plastic zone over a larger width. Furthermore, by comparing the snapshots on the left hand side of Fig.~\ref{fig: Tensiletest_plot} to their counterparts on the right hand side, one can conclude that parameter $c_1$ is large enough to approximate $\alpha^{(i)}=\varphi^{(i}$, i.e., the distributions of the accumulated plastic strains an those corresponding to the micromorphic field agree well (In order to distinguish both fields, a different color map has been used on purpose.). Since the novel update based on the augmented Lagrangian formulation leads to identical results as the novel variational update combined with the Fischer-Burmeister-function, the respective results are not explicitly shown here.

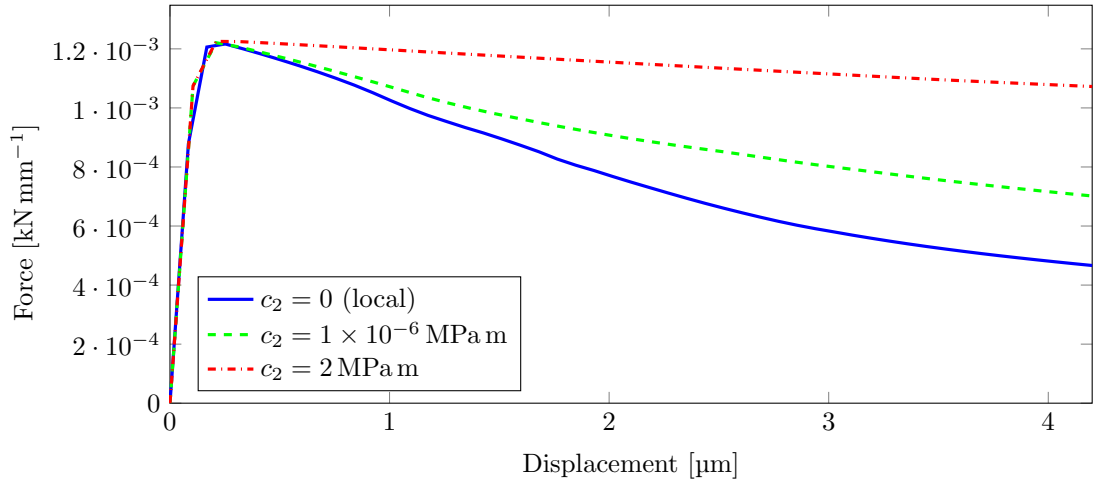
\begin{figure}[ht]
  \centering
  \begin{minipage}[t]{0.02\textwidth}
    \vspace{45pt}
    \centering
    \rotatebox{90}{Force [\si{\kilo\newton\per\milli\metre}]}
  \end{minipage}
  \begin{minipage}[t]{0.9\textwidth}
    \vspace{0pt}
    \centering
    \begin{tikzpicture}
      \begin{axis}[
        width = 1.\linewidth, 
        height = 0.5\linewidth,
        xmin =  0.0, 
        xmax =  4.2, 
        ymin =  0.0, 
        xtick = {0.0, 1.0, 2.0, 3.0, 4.0},
        x tick label style={/pgf/number format/fixed},
        scaled ticks=false, 
        xlabel = {Displacement [\si{\micro\metre}]},
        legend cell align=left,
        legend pos=south west
        ]
        \addplot[color = blue, 
                 mark = \empty,
                 line width = 1.2pt]
                 table {./Tensile_idPlast_Var_BE_local.txt};
        \addplot[color = green,
                 mark = \empty,
                 dashed,
                 line width = 1.2pt]
                 table {./Tensile_idPlast_Var_BE_1em6.txt};
        \addplot[color = red,
                mark = \empty,
                dash dot,
                line width = 1.2pt]
                table {./Tensile_idPlast_Var_BE_2.txt};
        \legend{$c_2=0$ (local),$ c_2 = \SI{1e-6}{\mega\pascal\metre} $ ,  $ c_2 = \SI{2}{\mega\pascal\metre} $}
      \end{axis}
      \end{tikzpicture}
  \end{minipage}
  \caption{Tensile test of a single crystal: Force-Displacement response for three different length scale parameters: $c_2=0$ (local model), $c_2=1\times 10^{-6}$~MPam and $c_2=2$~MPam; see Eq.~\eqref{eq-psi-micromorph}$_2$. Since plane strain conditions are considered, the force has been normalized by a thickness.}
  \label{fig: Tensiletest_forcedisp}
\end{figure}

Alternatively to analyzing the width of the plastic zone, the influence of gradient hardening can also be investigated by the resulting force displacement diagram, cf. Fig.~\ref{fig: Tensiletest_forcedisp}. 
As expected, the local model without hardening ($c_2=0$) leads to geometric softening. By increasing $c_2$, gradient hardening becomes active and consequently, the aforementioned softening is less pronounced. However, even choice $c_2=2$~MPam is not enough, to eliminate the overall softening completely.

\section{CONCLUSIONS}\label{sec:conclusions}

In this paper, two numerically robust algorithmic formulations suitable for rate-independent gradient-enhanced crystal plasticity were presented. Both of them rely on the underlying variational structure. The key ingredient of this structure is an incremental potential whose minimizers define every aspect of the model: the constitutive model as well as the balance equation. From a mathematical point of view, the computation of the unknowns follows from a constrained minimization problem. For this reason, powerful techniques from non-linear optimization can be applied. Two such techniques were elaborated and analyzed in this paper. 

The first algorithm is based on a nonlinear complementarity problem (NCP). To be more precise, a Fischer-Burmeister function was considered. It was shown that by introducing a new scaling parameter $w$ within a smoothed Fischer-Burmeister function, the resulting set of equations can be solved by a standard Newton method. Particularly, the tangent matrix within the Newton iteration is regular. By way of contrast it was also demonstrated that the non-smoothed Fischer-Burmeister function $(\delta=0)$ inherits the non-uniqueness of the original crystal plasticity model. It bears emphasis that these findings are not restricted to the considered framework of energy minimization, but also hold for conventional algorithmic formulations of crystal plasticity theory, e.g., for a return-mapping scheme combined with the Fischer-Burmeister function. With respect to conventional algorithmic formulations such as the return-mapping scheme coupled with an active-set strategy, it was also shown that they can be interpreted as a nonlinear complementarity problem (NCP) based on the Min-function.

The second algorithm is based on an augmented Lagrangian approach. Here, a direct incorporation of the admissibility condition of the stresses was not possible within the variational framework. For this reason, an alternative ansatz was proposed. Essentially, the yield function within a return-mapping-based algorithm was replaced by the necessary extremum condition of the incremental potential. Again, it was shown that tangent matrix associated with the resulting scheme is regular and thus, the solution of each Newton iteration is well-defined.

Both elaborated algorithmic formulations suitable for rate-independent gradient-enhanced crystal plasticity are robust, efficient and hence, indeed promising. Particularly, the tangent matrices corresponding to both schemes are regular. In future work, the efficiency of theses approaches is to be further improved by combing them --- also with sound active set methods.

\section*{Acknowledgments}
Financial support from the German Research Foundation (DFG) via SFB/TR TRR 188, project C04, is gratefully acknowledged.

\appendix
\section{Variational constitutive updates for orthogonal slip systems}\label{subsec:orthogonal-slip}

In general, variational constitutive updates such as those presented in this paper are not equivalent to classic return-mapping-type algorithms, cf. \cite{Mosler-Bleier2012}. In what follows, it is shown that this equivalence is fulfilled for orthogonal slip systems with property $ \vec{M}\ssi \perp \vec{N}\ssj \, \forall i,j \in \aset $. In this case, a straightforward calculation based on a Taylor series representation of the exponential map combined with initial condition $\mat{F}\pl(t_0)=\mat{1}$ gives
\eq{eq-EXPM-series-perpslips}{
	\mat{F}\pl=\EXP ( \Delta \Lp)\cdot\mat{F}\pl_{n-1} =\identity + \sum\limits_{t=t_0}^{t_n}\sum\limits_ {i=1}^{n_\text{sys}} \dlambda\ssi_t \vec{M}\ssi\otimes\vec{N}\ssi
}
The same equation is also fulfilled, if a backward-Euler integration is applied. According to Eq.~\eqref{eq-EXPM-series-perpslips},
\eq{eq-EXPM-series-perpslips-1-1}{
	\partial_{\dlambda\ssi_{t_n}} \F\pl =
	\vec{M}\ssi\otimes\vec{N}\ssi
}
and
\eq{eq-EXPM-series-perpslips-3}{
	{\mat{F}\pl}^{-1}=\identity - \sum\limits_{t=t_0}^{t_n}\sum\limits_ {i=1}^{n_\text{sys}} \dlambda\ssi_t \vec{M}\ssi\otimes\vec{N}\ssi
}
and thus
\eq{eq-EXPM-series-perpslips-4}{
	\partial_{\dlambda\ssi_{t_n}} \F\pl \cdot \inverse{ \left[ \F\pl \right]}=
	\left[\vec{M}\ssi\otimes\vec{N}\ssi\right]\cdot
}
By inserting equations \eqref{eq-EXPM-series-perpslips-4} into \eqref{eq-ex-dIdlamda}, one finally obtains identity
\eq{eq-ex-dIdlamda-special}{
	\dIdlambda = - \mat{\Sigma} : \underbrace{\left[ \partial_{\dlambda\ssi} \F\pl \cdot \inverse{ \left[ \F\pl \right]} \right]}_{ \vec{M}\ssi\otimes\vec{N}\ssi} + Q_0 + Q\ssi+c_1\,\left(\alpha^{(i)}-s^{(i)}\right)
	= - \phi^{(i),\text{nonl}}.
}
Consequently, a minimization of potential $\mathcal{I}_{\text{inc}}$ enforces the yield function exactly in the case of $ \vec{M}\ssi \perp \vec{N}\ssj \, \forall i,j \in \aset $.

\bibliographystyle{apalike-ejor}
\bibliography{./ABK.bib,./bibliography,./Quellen}

\begin{thebibliography}{}

\bibitem[Anand \& Kothari, 1996]{ANAND1996525}
Anand, L. \& Kothari, M. (1996).
\newblock A computational procedure for rate-independent crystal plasticity.
\newblock {\em Journal of the Mechanics and Physics of Solids}, 44(4),
  525--558.
\newblock \url{https://doi.org/https://doi.org/10.1016/0022-5096(96)00001-4}

\bibitem[Bassani \& Wu, 1991]{Bassani_Wu:1991}
Bassani, J. \& Wu, T.-Y. (1991).
\newblock Latent hardening in single crystals ii. analytical characterisation
  and prediction.
\newblock {\em Proceedingsof the Royal Society of London A.}, 435, 21--41.

\bibitem[Berdichevsky, 2006]{Berdichevsky_2006}
Berdichevsky, V.~L. (2006).
\newblock Continuum theory of dislocations revisited.
\newblock {\em Continuum Mechanics and Thermodynamics}, 18(3), 195.
\newblock \url{https://doi.org/10.1007/s00161-006-0024-7}

\bibitem[Bishop \& Hill, 1951a]{Bishop_Hill:51b}
Bishop, J. \& Hill, R. (1951a).
\newblock A theoretical derivation of the plastic properties of polycrystalline
  face-centered metals.
\newblock {\em Philos. Mag.}, 42, 1298--1307.

\bibitem[Bishop \& Hill, 1951b]{Bishop_Hill:51a}
Bishop, J. \& Hill, R. (1951b).
\newblock A theory of the plastic distortion of a polycrystal aggregate under
  combined stresses.
\newblock {\em Philos. Mag.}, 42, 414--417.

\bibitem[Bleier \& Mosler, 2012]{Mosler-Bleier2012}
Bleier, N. \& Mosler, J. (2012).
\newblock Efficient variational constitutive updates by means of a novel
  parameterization of the flow rule.
\newblock {\em International Journal for Numerical Methods in Engineering},
  89(9), 1120--1143.
\newblock \url{https://doi.org/10.1002/nme.3280}

\bibitem[Carstensen et~al., 2002]{Carstensen_Hackl_Mielke:02}
Carstensen, C., Hackl, K., \& Mielke, A. (2002).
\newblock Non-convex potentials and microstructures in finite-strain
  plasticity.
\newblock {\em {P}roc. {R}. {S}oc. {L}ond. {A}}, 458, 299--317.

\bibitem[Coleman \& Noll, 1963]{ColemanNoll1963}
Coleman, B. \& Noll, W. (1963).
\newblock The thermodynamics of elastic materials with heat conduction and
  viscosity.
\newblock {\em Archive for Rational Mechanics and Analysis}, 13(1), 167--178.
\newblock \url{https://doi.org/10.1007/BF01262690}

\bibitem[Cuitino \& Ortiz, 1993]{AMCuitino_1993}
Cuitino, A.~M. \& Ortiz, M. (1993).
\newblock Computational modelling of single crystals.
\newblock {\em Modelling and Simulation in Materials Science and Engineering},
  1(3), 225.
\newblock \url{https://doi.org/10.1088/0965-0393/1/3/001}

\bibitem[Daroju et~al., 2022]{Daroju2022}
Daroju, S., Kuwabara, T., Sharma, R., Fullwood, D.~T., Miles, M.~P., \&
  Knezevic, M. (2022).
\newblock {Experimental characterization and crystal plasticity modeling for
  predicting load reversals in AA6016-T4 and AA7021-T79}.
\newblock {\em International Journal of Plasticity}, 153, 103292.
\newblock \url{https://doi.org/10.1016/j.ijplas.2022.103292}

\bibitem[Fohrmeister et~al., 2018a]{FohrmeisterBartelsMosler_2018}
Fohrmeister, V., Bartels, A., \& Mosler, J. (2018a).
\newblock Variational updates for thermomechanically coupled gradient-enhanced
  elastoplasticity — implementation based on hyper-dual numbers.
\newblock {\em Computer Methods in Applied Mechanics and Engineering}, 339,
  239--261.
\newblock \url{https://doi.org/10.1016/j.cma.2018.04.047}

\bibitem[Fohrmeister et~al., 2018b]{FohrmeisterDiazMosler_2018a}
Fohrmeister, V., Díaz, G., \& Mosler, J. (2018b).
\newblock Classic crystal plasticity theory vs crystal plasticity theory based
  on strong discontinuities—theoretical and algorithmic aspects.
\newblock {\em International Journal for Numerical Methods in Engineering},
  0(0).
\newblock \url{https://doi.org/10.1002/nme.6000}

\bibitem[Forest, 2009]{Forest_2009}
Forest, S. (2009).
\newblock Micromorphic approach for gradient elasticity, viscoplasticity, and
  damage.
\newblock {\em Journal of Engineering Mechanics}, 135(3), 117--131.
\newblock \url{https://doi.org/10.1061/(ASCE)0733-9399(2009)135:3(117)}

\bibitem[Frank, 1951]{Frank1951}
Frank, F. (1951).
\newblock Lxxxiii. crystal dislocations.—elementary concepts and definitions.
\newblock {\em The London, Edinburgh, and Dublin Philosophical Magazine and
  Journal of Science}, 42(331), 809--819.
\newblock \url{https://doi.org/10.1080/14786445108561310}

\bibitem[Geiger \& Kanzow, 1999]{Geiger_Kanzow:02}
Geiger, C. \& Kanzow, C. (1999).
\newblock {\em Theorie und {N}umerik restringierter {O}ptimierungsaufgaben}.
\newblock Springer.

\bibitem[Gurtin, 2006]{Gurtin:06}
Gurtin, M. (2006).
\newblock The burgers vector and the ﬂow of screw and edge dislocations in
  ﬁnite-deformation single-crystal plasticity.
\newblock {\em Journal of the Mechanics and Physics of Solids}, 54, 1882--1898.

\bibitem[Gurtin, 1996]{Gurtin1996}
Gurtin, M.~E. (1996).
\newblock {Generalized Ginzburg-Landau and Cahn-Hilliard equations based on a
  microforce balance}.
\newblock {\em Physica D: Nonlinear Phenomena}, 92(3--4), 178 -- 192.
\newblock \url{https://doi.org/10.1016/0167-2789(95)00173-5}

\bibitem[Havner, 1992]{Havner:1992}
Havner, K. (1992).
\newblock {\em Finite Deformation of Crystalline Solids}.
\newblock Cambridge University Press.

\bibitem[Hill \& Rice, 1972]{Hill_Rice:1972}
Hill, R. \& Rice, J. (1972).
\newblock Constitutive analysis of elastic-plastic crystals at arbitrary
  strain.
\newblock {\em Journal of the Mechanics and Physics of Solids}, 20, 401--413.

\bibitem[Hollenweger \& Kochmann, 2022]{Kochmann2022}
Hollenweger, Y. \& Kochmann, D.~M. (2022).
\newblock {An efficient temperature-dependent crystal plasticity framework for
  pure magnesium with emphasis on the competition between slip and twinning}.
\newblock {\em International Journal of Plasticity}, 158, 103448.
\newblock \url{https://doi.org/10.1016/j.ijplas.2022.103448}

\bibitem[Homayonifar \& Mosler, 2011]{Homayonifar2011}
Homayonifar, M. \& Mosler, J. (2011).
\newblock {On the coupling of plastic slip and deformation-induced twinning in
  magnesium: A variationally consistent approach based on energy minimization}.
\newblock {\em International Journal of Plasticity}, 27(7), 983--1003.
\newblock \url{https://doi.org/10.1016/j.ijplas.2010.10.009}

\bibitem[Homayonifar \& Mosler, 2012]{Homayonifar2012}
Homayonifar, M. \& Mosler, J. (2012).
\newblock {Efficient modeling of microstructure evolution in magnesium by
  energy minimization}.
\newblock {\em International Journal of Plasticity}, 28(1), 1--20.
\newblock \url{https://doi.org/10.1016/j.ijplas.2011.05.011}

\bibitem[Langenfeld et~al., 2018]{LangenfeldJunkerMosler_2018}
Langenfeld, K., Junker, P., \& Mosler, J. (2018).
\newblock Quasi-brittle damage modeling based on incremental energy relaxation
  combined with a viscous-type regularization.
\newblock {\em Continuum Mechanics and Thermodynamics}, 30(5), 1125--1144.
\newblock \url{https://doi.org/10.1007/s00161-018-0669-z}

\bibitem[Le \& G\"unther, 2014]{Le2014}
Le, K. \& G\"unther, C. (2014).
\newblock Nonlinear continuum dislocation theory revisited.
\newblock {\em International Journal of Plasticity}, 53, 164--178.
\newblock \url{https://doi.org/10.1016/j.ijplas.2013.08.003}

\bibitem[{Lee}, 1969]{Lee1969}
{Lee}, E.~H. (1969).
\newblock {Elastic-Plastic Deformation at Finite Strains}.
\newblock {\em Journal of Applied Mechanics}, 36, 1--6.
\newblock \url{https://doi.org/10.1115/1.3564580}

\bibitem[Ma et~al., 2006]{MA20062181}
Ma, A., Roters, F., \& Raabe, D. (2006).
\newblock On the consideration of interactions between dislocations and grain
  boundaries in crystal plasticity finite element modeling – theory,
  experiments, and simulations.
\newblock {\em Acta Materialia}, 54(8), 2181--2194.
\newblock \url{https://doi.org/https://doi.org/10.1016/j.actamat.2006.01.004}

\bibitem[Manik et~al., 2022]{Manik_Asadkandi_Holmedal:2022}
Manik, T., Asadkandi, H., \& Holmedal, B. (2022).
\newblock A robust algorithm for rate-independent crystal plasticity.
\newblock {\em Computer Methods in Applied Mechanics and Engineering}, 393,
  114831.

\bibitem[M\'{a}nik \& Holmedal, 2014]{Manik2014}
M\'{a}nik, T. \& Holmedal, B. (2014).
\newblock Review of the taylor ambiguity and the relationship between
  rate-independent and rate-dependent full-constraints taylor models.
\newblock {\em International Journal of Plasticity}, 55, 152--181.
\newblock \url{https://doi.org/10.1016/j.ijplas.2013.10.002}

\bibitem[Miehe, 2002]{Miehe:02}
Miehe, C. (2002).
\newblock Strain-driven homogenization of inelastic microstructures and
  composites based on an incremental variational formulation.
\newblock {\em International Journal for Numerical Methods in Engineering}, 55,
  1285--1322.

\bibitem[Miehe et~al., 2002]{Miehe2002}
Miehe, C., Schotte, J., \& Lambrecht, M. (2002).
\newblock Homogenization of inelastic solid materials at finite strains based
  on incremental minimization principles. application to the texture analysis
  of polycrystals.
\newblock {\em Journal of the Mechanics and Physics of Solids}, 50(10), 2123 --
  2167.
\newblock \url{https://doi.org/10.1016/S0022-5096(02)00016-9}

\bibitem[Miehe \& Schröder, 2001]{MieheSchroeder_2001}
Miehe, C. \& Schröder, J. (2001).
\newblock A comparative study of stress update algorithms for rate-independent
  and rate-dependent crystal plasticity.
\newblock {\em International Journal for Numerical Methods in Engineering},
  50(2), 273--298.
\newblock
  \url{https://doi.org/10.1002/1097-0207(20010120)50:2<273::AID-NME17>3.0.CO;2-Q}

\bibitem[Mosler \& Bruhns, 2010]{Mosler-Bruhns2010}
Mosler, J. \& Bruhns, O. (2010).
\newblock On the implementation of rate-independent standard dissipative solids
  at finite strain – variational constitutive updates.
\newblock {\em Computer Methods in Applied Mechanics and Engineering},
  199(9–12), 417--429.
\newblock \url{https://doi.org/10.1016/j.cma.2009.07.006}

\bibitem[Nieh\"user \& Mosler, 2023]{Niehueser_Mosler:23}
Nieh\"user, A. \& Mosler, J. (2023).
\newblock Numerically efficient and robust interior-point algorithm for finite
  strain rate-independent crystal plasticity.
\newblock {\em Computer Methods in Applied Mechanics and Engineering}.
\newblock in press.

\bibitem[Nye, 1953]{NYE1953153}
Nye, J. (1953).
\newblock Some geometrical relations in dislocated crystals.
\newblock {\em Acta Metallurgica}, 1(2), 153--162.
\newblock \url{https://doi.org/https://doi.org/10.1016/0001-6160(53)90054-6}

\bibitem[Ortiz \& Repetto, 1999]{Ortiz_Repetto:99}
Ortiz, M. \& Repetto, E. (1999).
\newblock Nonconvex energy minimisation and dislocation in ductile single
  crystals.
\newblock {\em {J}. {M}ech. {P}hys. {S}olids}, 47, 397--462.

\bibitem[Ortiz \& Stainier, 1999]{Ortiz_Stainier:99}
Ortiz, M. \& Stainier, L. (1999).
\newblock The variational formulation of viscoplastic constitutive updates.
\newblock {\em Computer Methods in Applied Mechanics and Engineering}, 171,
  419--444.

\bibitem[Pr{\"u}ger \& Kiefer, 2020]{Prger2020ACS}
Pr{\"u}ger, S. \& Kiefer, B. (2020).
\newblock A comparative study of integration algorithms for finite single
  crystal (visco-)plasticity.
\newblock {\em International Journal of Mechanical Sciences}, 180, 105740.

\bibitem[Ry\'{s} et~al., 2020]{Rys2022b}
Ry\'{s}, M., Forest, S., \& Petryk, H. (2020).
\newblock A micromorphic crystal plasticity model with the gradient-enhanced
  incremental hardening law.
\newblock {\em International Journal of Plasticity}, 128, 102655.
\newblock \url{https://doi.org/10.1016/j.ijplas.2019.102655}

\bibitem[Ry\'{s} et~al., 2022]{Rys2022}
Ry\'{s}, M., Stupkiewicz, S., \& Petryk, H. (2022).
\newblock Micropolar regularization of crystal plasticity with the
  gradient-enhanced incremental hardening law.
\newblock {\em International Journal of Plasticity}, 156, 103355.
\newblock \url{https://doi.org/10.1016/j.ijplas.2022.103355}

\bibitem[Scheunemann et~al., 2021]{SCHEUNEMANN2021111149}
Scheunemann, L., Nigro, P., \& Schröder, J. (2021).
\newblock Numerical treatment of small strain single crystal plasticity based
  on the infeasible primal-dual interior point method.
\newblock {\em International Journal of Solids and Structures}, 232, 111149.
\newblock \url{https://doi.org/https://doi.org/10.1016/j.ijsolstr.2021.111149}

\bibitem[Scheunemann et~al., 2020]{SCHEUNEMANN20201}
Scheunemann, L., Nigro, P., Schröder, J., \& Pimenta, P. (2020).
\newblock A novel algorithm for rate independent small strain crystal
  plasticity based on the infeasible primal-dual interior point method.
\newblock {\em International Journal of Plasticity}, 124, 1--19.
\newblock \url{https://doi.org/https://doi.org/10.1016/j.ijplas.2019.07.020}

\bibitem[Schmidt-Baldassari, 2003]{Schmidt-Baldassari:03}
Schmidt-Baldassari, M. (2003).
\newblock Numerical concepts for rate-independent single crystal plasticity.
\newblock {\em Computer Methods in Applied Mechanics and Engineering}, 192,
  1261--1280.

\bibitem[Schröder \& Miehe, 1997]{SCHRODER1997168}
Schröder, J. \& Miehe, C. (1997).
\newblock Aspects of computational rate-independent crystal plasticity.
\newblock {\em Computational Materials Science}, 9(1), 168--176.
\newblock \url{https://doi.org/https://doi.org/10.1016/S0927-0256(97)00072-4}.
\newblock Selected papers of the Sixth International Workshop on Computational
  Mechanics of Materials

\bibitem[Simo, 1998]{Simo1998}
Simo, J. (1998).
\newblock Numerical analysis and simulation of plasticity.
\newblock {\em Handbook of Numerical Analysis}, volume~6 of {\em Handbook of
  Numerical Analysis}, 183--499. Elsevier.
\newblock \url{https://doi.org/10.1016/S1570-8659(98)80009-4}

\bibitem[Simo \& Hughes, 1998]{SimoHughes1998}
Simo, J. \& Hughes, T. (1998).
\newblock {\em Computational inelasticity}, volume~7 of {\em Interdisciplinary
  Applied Mathematics}.
\newblock Springer-Verlag, New York.

\bibitem[Stainier \& Ortiz, 2010]{Stainier2010}
Stainier, L. \& Ortiz, M. (2010).
\newblock Study and validation of a variational theory of thermo-mechanical
  coupling in finite visco-plasticity.
\newblock {\em International Journal of Solids and Structures}, 47(5),
  705--715.
\newblock \url{https://doi.org/10.1016/j.ijsolstr.2009.11.012}

\bibitem[Steinmann, 2015]{Steinmann2015-js}
Steinmann, P. (2015).
\newblock {\em Geometrical foundations of continuum mechanics} (2015 ed.).
\newblock Lecture Notes in Applied Mathematics and Mechanics. Springer.
\newblock \url{https://doi.org/https://doi.org/10.1007/978-3-662-46460-1}

\bibitem[Taylor, 1938]{Taylor1938}
Taylor, G.~I. (1938).
\newblock Plastic strain in metals.
\newblock {\em Journal of the Institute of Metals}, 62, 307--324.

\bibitem[Zuo, 2011]{Zuo2012}
Zuo, Q. (2011).
\newblock On the uniqueness of a rate-independent plasticity model for single
  crystals.
\newblock {\em International Journal of Plasticity}, 27(8), 1145--1164.
\newblock \url{https://doi.org/10.1016/j.ijplas.2010.12.002}

\bibitem[Šiška et~al., 2009]{SISKA2009793}
Šiška, F., Weygand, D., Forest, S., \& Gumbsch, P. (2009).
\newblock Comparison of mechanical behaviour of thin film simulated by discrete
  dislocation dynamics and continuum crystal plasticity.
\newblock {\em Computational Materials Science}, 45(3), 793--799.
\newblock
  \url{https://doi.org/https://doi.org/10.1016/j.commatsci.2008.07.006}.
\newblock Proceedings of the 17th International Workshop on Computational
  Mechanics of Materials

\end{thebibliography}

\end{document}